\documentclass[12pt]{article}
\usepackage{amsmath,amssymb,epsfig,amsfonts}
\usepackage{graphicx,subfigure}
\usepackage{setspace}
\usepackage{cite}
\usepackage[usenames,dvipsnames]{xcolor}
\usepackage{hyperref}
\usepackage{tikz}
\usepackage{lscape}
\hypersetup{urlcolor=blue}
\usetikzlibrary{decorations.markings}
\tikzstyle{Vertex}=[circle, draw, inner sep=0pt, minimum size=6pt]

\usepackage[top=1.25in,bottom=1.25in,right=0.95in,left=0.95in]{geometry}

\makeatletter

\newcommand{\pqvec}[2]{\scriptsize \left( \begin{array}{c}
#1 \\
#2
\end{array} \right)}

\newcommand{\bZ}{\mathbb{Z}}
\newcommand{\bC}{\mathbb{C}}
\newcommand{\bP}{\mathbb{P}}
\newcommand{\bR}{\mathbb{R}}
\newcommand{\bF}{\mathbb{F}}

\newcommand{\cN}{\mathcal{N}}

\newcommand{\cO}{\mathcal{O}}

\newcommand{\ep}{\varepsilon}
\newcommand{\wgdef}[1]{(f,g)\mapsto (f,g+#1)}

\newcommand{\wdef}[2]{(f,g)\mapsto (f+#1,g+#2)}

\def\unit{{1\kern-.65ex {\rm l}}}
\def\1{{1\kern-.65ex {\rm l}}}

\definecolor{xdotcolor}{named}{ForestGreen}
\definecolor{tmotioncolor}{named}{Cerulean}

\newcount\hour \newcount\minute
\hour=\time \divide \hour by 60
\minute=\time
\def\now{%
\ifnum \hour<13
  \ifnum \hour=0 \advance \hour by 12 \number\hour:\else \number\hour:\fi%
     \ifnum \minute<10 0\fi%
     \number\minute%
\ A.M.%
\else \advance \hour by -12 \number\hour:%
  \ifnum \minute<10 0\fi%
  \number\minute%
  \ P.M.%
\fi%
}

\makeatother

\begin{document}

% format
\baselineskip=18pt  % a la harvmac
\numberwithin{equation}{section}  % make eq labels (sec.num)
\allowdisplaybreaks  % allow page breaks in displayed eqs

% print date, time and filename
%\pagestyle{myheadings}
%\markright{{\tt \jobname.tex} -- \today{} \now}

%%%%%%%%%%%%%%%%%%%%%%%%%%%%%%%%%%%%%%%%%%%
%%%        TITLE BEGINS HERE
%%%%%%%%%%%%%%%%%%%%%%%%%%%%%%%%%%%%%%%%%%%

%% ========== title (note version) begins here ==========
%
%\vspace*{-1cm}
%\begin{center}
% {\Large\bf Title of the Document}
%\end{center}
%\vspace*{-.5cm}
%
%% ========== title (note version) ends here ==========

%% ========== title (paper version, a la harvmac) begins here ==========

\thispagestyle{empty}

% Report number
\vspace*{-2cm}
\begin{flushright}
{\tt NSF-KITP-14-010}
\end{flushright}

% title, authors, affiliation
\vspace*{3.8cm}
\begin{center}
 {\LARGE Non-Abelian Gauge Symmetry \\ \vspace{.3cm} and the Higgs Mechanism in F-theory}
 \vspace*{1.0cm}

Antonella Grassi$^1$, James Halverson$^2$, and Julius L. Shaneson$^1$

\vspace{1.0cm}

$^1$ Department of Mathematics, University of Pennsylvania\\
Philadelphia, PA 19104-6395 USA \\
$^2$ Kavli Institute for Theoretical Physics, University of California\\
Santa Barbara, CA 93106-4030 USA \\

\end{center}
\vspace*{1cm}

\abstract{ \vspace{.5cm}Singular fiber resolution does not describe
  the spontaneous breaking of gauge symmetry in F-theory, as the
  corresponding branch of the moduli space does not exist in the
  theory.  Accordingly, even non-abelian gauge theories have not been
  fully understood in global F-theory compactifications. We present a
  systematic discussion of using singularity deformation, which does
  describe the spontaneous breaking of gauge symmetry in F-theory, to
  study non-abelian gauge symmetry.  Since this branch of the moduli
  space also exists in the defining M-theory compactification, it
  provides the only known description of gauge theory states which
  exists in both pictures; they are string junctions in F-theory. We
  discuss how global deformations give rise to local deformations, and
  also give examples where local deformation can be utilized even in
  models where a global deformation does not exist.  Utilizing
  deformations, we study a number of new examples, including
  non-perturbative descriptions of $SU(3)$ and $SU(2)$ gauge theories
  on seven-branes which do not admit a weakly coupled type IIb
  description.  It may be of phenomenological interest that these
  non-perturbative descriptions do not exist for higher rank $SU(N)$
  theories.  }

\clearpage

\onehalfspacing

\section{Introduction}
\label{sec:introduction}

Much has been learned about the landscape of string vacua over the
last fifteen years. On one hand, in weakly coupled corners of the
landscape there are scenarios for controlled moduli stabilization
which are giving rise to increasingly realistic global
compactifications, and in many cases provide inspiration for new
models of particle physics or early universe cosmology. On the other
hand, there has been much formal progress in understanding the physics
of compactifications at small volume or strong coupling. We have
gained both a clearer view of well-known regions, and also a glimpse
of relatively unexplored vistas.

The strongest example of the latter may be in compactifications of
F-theory\cite{Vafa:1996xn}, which has been the subject of much study
in recent years.  In addition to enjoying
\cite{Beasley:2008dc,Donagi:2008ca} certain model-building advantages
for grand unification over their heterotic and type II string
counterparts, F-theory compactifications provide a broad view of the
landscape; much of their strongly coupled physics can be understood in
terms of the geometry of elliptically fibered Calabi-Yau varieties,
which can be explicitly constructed and studied as a function of their
moduli.  Recent works have made significant progress in understanding
the physics of F-theory compactifications, including globally
consistent models \cite{Andreas:2009uf, Marsano:2009ym,
  Collinucci:2009uh, Blumenhagen:2009up, Marsano:2009gv,
  Blumenhagen:2009yv, Marsano:2009wr, Grimm:2009yu, Cvetic:2010rq,
  Chen:2010ts, Chen:2010tp, Chung:2010bn,
  Chen:2010tg,Knapp:2011wk,Knapp:2011ip,Marsano:2012yc}, $U(1)$
symmetries \cite{Grimm:2010ez, Dolan:2011iu, Marsano:2011nn,
  Grimm:2011tb,
  Morrison:2012ei,Mayrhofer:2012zy,Cvetic:2013nia,Cvetic:2013jta,Cvetic:2013jta,Cvetic:2013qsa,Borchmann:2013jwa,Borchmann:2013hta,Braun:2013nqa,Braun:2013yti,Braun:2014nva},
instanton corrections \cite{Blumenhagen:2010ja,Cvetic:2010rq,
  Donagi:2010pd,
  Grimm:2011dj,Marsano:2011nn,Cvetic:2011gp,Bianchi:2011qh,Kerstan:2012cy,Cvetic:2012ts,Bianchi:2012kt},
the physics of codimension two and three singularities
\cite{Grassi:2000we,Grassi:2011hq,Morrison:2011mb,Grassi:2013kha,Lawrie:2012gg,Hayashi:2013lra,Hayashi:2014kca,Bonora:2010bu,Mizoguchi:2014gva},
and chirality inducing $G_4$-flux \cite{Marsano:2010ix,
  Collinucci:2010gz,Marsano:2011nn, Braun:2011zm, Marsano:2011hv,
  Krause:2011xj, Grimm:2011fx, Braun:2012nk,
  Kuntzler:2012bu,Lawrie:2012gg, Krause:2012he,
  Collinucci:2012as,Bianchi:2012kt,Marsano:2012bf,Intriligator:2012ue,Anderson:2013rka
}; there has also been progress in understanding the landscape of
six-dimensional F-theory compactifications
\cite{Kumar:2009ac,Kumar:2010ru,Kumar:2010am,Morrison:2012np,Morrison:2012ei}.

In this paper we study non-abelian gauge symmetry as it exists in
global F-theory compactifications. There, non-abelian sectors (not
arising from D3 branes) require a singular compactification geometry
$X$. Lacking both the tools to deal directly with the singular
geometry and also a fundamental quantization of M-theory, which
defines any F-theory compactification, one must resort to studying the
theory on a related smooth manifold $\tilde X$ and then inferring the
physics as one takes the singular limit $\tilde X \rightarrow
X$. Since a singular geometry is necessary for a non-abelian gauge
sector and $\tilde X$ is smooth, the movement in moduli space
$X\rightarrow \tilde X$ must describe spontaneous symmetry breaking
via the Higgs mechanism. The Lie algebraic data of particle states in
the broken gauge theory is encoded in the geometry of $\tilde X$.

As $X$ is a Calabi-Yau variety, the smoothing processes necessarily
correspond to movement in the K\" ahler or complex structure moduli
spaces of $X$. A common technique is to study F-theory via M-theory on
a related singular variety $X_M$, where $X_M\rightarrow X$ is the
limit of vanishing elliptic fiber, and then to blow-up $X_M$. This
branch of the moduli space is often referred to as the M-theory
Coulomb branch (which is a bit of a misnomer), but it does
not exist in F-theory on $X$; as such, it does not describe the
spontaneous breaking of \emph{any} non-abelian gauge symmetry in
F-theory. Instead, the breaking of gauge symmetry in F-theory is
accomplished by singularity deformation via movement in complex
structure. This description of spontaneous symmetry breaking exists
both for M-theory on $X_M$ and F-theory on $X$.

This paper is organized as follows. First we will review F-theory, as
well as the drawbacks of resolution and the advantages of deformation
as a technique for its study. In section \ref{sec:glimpse} we present
a simple example which demonstrates how the Higgs mechanism operates
in global F-theory compactifications. In section
\ref{sec:deformations} we present a general discussion of complex
structure deformations useful for studying F-theory, which applies
even when the singularities cannot be globally Higgsed, as in the case
of non-Higgsable clusters. In section \ref{sec:less studied fibers} we
present new realizations of $\mathfrak{su}(3)$ and $\mathfrak{su}(2)$ gauge algebras on
stacks of four and three seven branes; interestingly, these
descriptions do not exist in the type IIb limit or for higher rank $\mathfrak{su}(N)$
algebras.  Links to codes used to perform the computations in this
paper can be found in appendix \ref{sec:code}.

\noindent \emph{\textbf{F-theory and its defining M-theory
    compactification.}}

\vspace{.1cm}
\noindent In general, an F-theory compactification to $d$ dimensions
on an elliptically fibered Calabi-Yau variety $X$ is \emph{defined} to
be the vanishing fiber limit of an M-theory compactification to $d-1$
dimensions on an elliptically\footnote{Actually, only a genus-one
  fibration is required, as explored recently in
  \cite{Braun:2014oya}.} fibered Calabi-Yau variety $X_M$; for us this
will be a Weierstrass model. In taking this limit $X_M \rightarrow X$,
one of the dimensions decompactifies (via a fiberwise T-duality),
yielding a $d$-dimensional theory. This M-theory compactification is
often called the ``defining M-theory compactification,'' and a common
technique is to study F-theory on $X$ in terms of the defining
M-theory compactification on $X_M$.  Via circle compactification, both
the $d^{th}$ component of a gauge field and also scalars in F-theory
are scalars in the defining M-theory compactification; as such, there
are more scalars which can receive expectation values in the M-theory
compactification, and thus perhaps more branches of gauge theory
moduli space.

The non-abelian gauge structure of these compactifications
is determined by the singular geometry of $X$ and $X_M$. In
particular, if $X$ and $X_M$ exhibit singularities along a codimension
one locus $Z$ in the base $B$, both the F-theory and defining M-theory
compactifications enjoy a non-abelian gauge sector along $Z$. If the
metric moduli of $X$ or $X_M$ are then continuously varied such that
the singular locus becomes smooth, the respective non-abelian gauge
theory has been spontaneously broken, and the now smooth geometry
encodes features of the broken theory.  Since any such smoothing will
break the non-abelian gauge theory, it is natural in both M-theory on
$X_M$ and its limit of F-theory on $X$ to study all possible
smoothings of $X_M$ and $X$, and also the associated physics.

\vspace{.5cm}
\noindent\emph{The M-theory Coulomb branch of singularity resolution.}

One such smoothing is singularity resolution via blow-up; this is a
classical technique in algebraic geometry which can be used to resolve
singularities in $X_M$.  From the point of view of resolution, $X_M$
is on the boundary of a K\" ahler cone of a family of smooth
Calabi-Yau manifolds, and the resolution procedure simply involves
moving to the interior of the K\"ahler cone. For the singularities
which are typically\footnote{We mean the resolution of
  singular fibers. See \cite{Heckman:2013pva}, which uses
  resolution of curves in the base to study $d=6$ $(1,0)$ SCFTs.}
resolved in the defining M-theory compactification on $X_M$, this
involves movement in a direction in K\" ahler moduli space which cause
rational curves to appear in the fiber. For a generic such direction,
the gauge group $G$ of the theory along $Z$ is broken to
$U(1)^{rk(G)}$. This is one Coulomb branch, of perhaps many, of this
gauge theory in the defining M-theory compactification; henceforth, in
any reference to ``the'' M-theory Coulomb branch, we will mean the
M-theory Coulomb branch obtained by the resolution of singular
elliptic fibers. Gauge theoretically, this corresponds to giving
expectation values to the scalars in the $d-1$ dimensional theory
which are obtained from the $d$-dimensional theory via reduction of
the $d^{th}$ component of the gauge field. Alternatively, expectation
values of adjoint scalars in $d-1$ which are also adjoint scalars in
$d$ dimensions can give rise to Coulomb branches; see section
\ref{sec:deformation physics}. Singularity resolution can be a
useful tool because properties of the broken gauge theory are encoded
in the geometry of the resolved manifold.

However, there are important drawbacks to using the M-theory Coulomb branch
as a tool to study F-theory compactifications. These include:
\begin{itemize}
\item \emph{It doesn't exist in F-theory}. The scalars which have
  expectation values on the M-theory Coulomb branch become components
  of gauge fields in F-theory, and thus cannot receive expectation
  values in the F-theory limit. The M-theory Coulomb branch does not
  lift to F-theory. Therefore, it can describe neither the breaking of
  gauge symmetry nor massive gauge bosons in F-theory.

\item \emph{It is generically obstructed by instantons.} The Coulomb
  branch of $d=3$ $\cN=2$ gauge theories is typically lifted by
  instanton corrections; see \cite{Affleck:1982as} for early work and
  \cite{Aharony:1997bx,deBoer:1997kr} for further analysis. In
  M-theory compactifications these instanton corrections arise from
  $M5$-branes wrapped on Cartan divisors of the resolved geometry, as
  studied, for example, in \cite{Katz:1996th}. Physically, these
  effects imply that far out on the Coulomb branch there is a scalar
  potential obstructing the path back to the origin. To our knowledge,
  most recent works on the Coulomb branch do not take these effects
  into account.
\item \emph{It doesn't encompass the heterotic or type IIb
    perspective.} Under certain circumstances, an F-theory
  compactification can be dual to a heterotic compactification
  \cite{Vafa:1996xn,Morrison:1996na,Morrison:1996pp} or admit a weakly
  coupled type IIb description \cite{Vafa:1996xn,Sen:1996vd}. If so,
  the moduli which spontaneously break the heterotic or type IIb
  theories become complex structure moduli in F-theory. Thus, the
  M-theory Coulomb branch, on which K\" ahler moduli have expectation
  values, does not encompass the heterotic or type IIb perspectives.
\end{itemize}
Due to these drawbacks, it is natural to expect that utilizing the
M-theory Coulomb branch to study F-theory compactifications will miss
important aspects of F-theoretic physics; as we have mentioned, one
important omission is that it cannot describe massive gauge bosons in
F-theory. Accordingly, it is worthwhile to examine other possible
approaches.

\vspace{.5cm}
\noindent\emph{F-theory, M-theory, and Singularity Deformation}

We will take a different approach in this work. As any smoothing of
$X_M$ or $X$ will spontaneously break a non-abelian gauge theory in
the respective M-theory and F-theory compactifications, the other
natural choice is to break the theory by complex structure
deformation. (See \cite{Intriligator:2012ue} for a study
of resolution and deformation in the $d=3$ $\cN=2$ compactifications
of M-theory with  $G_4$ fluxes.)

Our method has the advantage that the corresponding branch of the moduli
space exists both in an F-theory compactification and its defining
M-theory compactification. In the $d=4$ case, this is simply the gauge
theoretic statement that the scalars in four-dimensional chiral
multiplets\footnote{When we
    discuss four-dimensional $\cN=1$ theories we refer to the
    supersymmetric multiplets with scalar fields as chiral multiplets,
    even though we do not address $G$-flux and therefore each chiral
    multiplet comes in a vector pair. This matches the standard
    language of $\cN=1$ theories in four dimensions even in the case
    of a non-chiral spectrum; see e.g. the classic paper on
    non-perturbative effects in supersymmetric QCD
    \cite{Affleck:1983mk}, which is a non-chiral theory. } can
receive expectation values in \emph{both} the four-dimensional theory
\emph{and} its circle compactification to three dimensions, in
contrast to the scalars of the M-theory Coulomb branch which can
receive expectations values only in three dimensions. Moreover,
studying F-theory via complex structure deformation has the advantage
that it encompasses and also extends, see section \ref{sec:less
  studied fibers},  some of the heterotic and type
IIb perspectives , since deformations of seven-branes in F-theory
  via complex structure deformation map to vector bundle and D7-brane
  moduli spaces in the heterotic and type IIb strings,
  respectively. 

We restrict our attention to utilizing singularity deformation to
study the structure of non-abelian gauge theories in F-theory. On this
branch of the moduli space in the M-theory description, the massive
states of a spontaneously broken non-abelian gauge theory are
described by $M2$-branes wrapped on two-cycles in the total space of
$X_M$ which extend exactly one dimension in the elliptic fiber. In the
F-theory limit $X_M\rightarrow X$, these states become the $(p,q)$
string junctions of
\cite{Gaberdiel:1997ud,DeWolfe:1998zf,Mikhailov:1998bx}; the Lie
algebraic structure of string junctions for certain sets of
vanishing cycles was pioneered in the work of Zwiebach
and DeWolfe \cite{DeWolfe:1998zf}.

The systematic study of such $M2$-brane states and string junctions in
terms of the deformation theory of elliptically fibered Calabi-Yau
varieties was initiated in our previous work \cite{Grassi:2013kha}. We
continue this program here and will also pursue it in forthcoming
works
\cite{GrassiHalversonShaneson:Math,GrassiHalversonShaneson:PhysicsCodim2}.
Thus, for the sake of brevity, we will only refer to the main concepts
of \cite{Grassi:2013kha} as they appear here, instead referring the
reader to our previous work for a systematic treatment of those
concepts.

\section{A Glimpse of the Higgs Mechanism in F-theory}
\label{sec:glimpse}

Before proceeding to a systematic discussion of deformations which is
sufficient to determine the Lie algebraic data of a spontaneously
broken gauge theory in F-theory, let us present a simple global
example which will give a useful and intuitive visual picture.  As a
brief caveat, in this section we will use language and techniques
which are less common in the F-theory literature.  We will attempt to
describe the salient points briefly, but refer the reader who is
interested in further details to our previous work
\cite{Grassi:2013kha}.

Consider an F-theory compactification to six dimensions on a Calabi-Yau elliptic fibration $\pi:\,X \rightarrow B$
with $B=\bP^2$ having homogeneous coordinates $(z,t,s)$. We will
consider a Weierstrass model defined by
\begin{equation}
  y^2 = x^3 + f\, x + g
\end{equation}
with $f$ and $g$ global sections of $\cO(12)$ and $\cO(18)$, respectively.
For generic $f$ and $g$ the manifold is smooth and the associated F-theory model has
no non-abelian gauge symmetry, which requires the presence of a singular codimension
one locus $Z$ in $B$.

Now we would like to engineer a non-abelian gauge theory via tuning
the complex structure moduli in $f$ and $g$. Suppose they take the
form $f=z^2\, p_{10}$ and $g=z^3\, p_{15}$ for general polynomials
$p_i$ of degree $i$ in the homogeneous coordinates, and furthermore
that $z$ divides neither $p_{10}$ nor $p_{15}$. Then the discriminant
$\Delta = 4\, f^3 + 27\, g^2$ for this example is given by
\begin{equation}
\Delta = z^6 \, \left(4 p_{10}^3+27 p_{15}^2\right) \equiv z^6\, \Delta_r
\end{equation}
and the theory exhibits an $I_0^*$ singularity along $Z\equiv
\{z=0\}$, a genus $0$ curve in $\bP^2$.  The curve $z=0$ intersect the
remaining discriminant in $30$ points, with transversal intersection,
as in {\cite{Grassi:2013kha}. This is a consistent global
  six-dimensional F-theory compactification with\footnote{If $p_{10}$
    and $p_{15}$ takes special forms the gauge symmetry may be
    $\mathfrak{so}(7)$ or $\mathfrak{so}(8)$; here it is
    $\mathfrak{g}_2$.}  $\mathfrak{g}_2$ gauge
  symmetry\footnote{Henceforth we will use algebra rather than
    notation, since we are only studying the Lie algebraic structure
    in the geometry.}.

We would like to study the Lie algebraic structure of the gauge theory
by performing a complex structure deformation which spontaneously
breaks the gauge theory. To do this, we consider the deformation
\begin{equation}
\wgdef{\epsilon\, p_{18}}
\end{equation}
where the degree $18$ polynomial $p_{18}$ is chosen such that $z$
does not divide it and $\ep \in \bC^*$. The deformed discriminant 
\begin{equation}
  \Delta = z^6\, \Delta_r
  +\epsilon \,(54 \,p_{15} \,p_{18} \, z^3 +27\, \epsilon\, p_{18}^2) \equiv z^6\, \Delta_r + \epsilon \, \Delta_\epsilon
\end{equation}
 is an $I_1$ locus. The non-abelian gauge theory
has been spontaneously broken. At generic points in neighborhood of
$z=0$, we have $\Delta_r\ne 0$ and $\Delta_\epsilon \ne 0$ and one can
study an elliptic fibration over the complex $z$-plane in this
neighborhood; in fact this is a family of elliptic surfaces,
where the parameters are the coordinates in the
base other than $z$.

In a generic elliptic surface in this family, prior to deformation,
the intersection of the discriminant with the $z$-plane gives a marked
point at $z=0$ with multiplicity $6$, which now has blossomed out to
give $6$ marked points with multiplicity $1$ collected around the
origin, and whose distance from the origin is set by $\epsilon\,
\Delta_\epsilon / \Delta_r$. This is the geometry transverse to the
Higgsed seven-brane for a generic point in its worldvolume, and the
Lie algebraic data of the Higgsed $\mathfrak{g}_2$ gauge theory must be
captured by the geometry. One can study the Lie algebraic structure of
two-cycles in a family of elliptic surfaces to recover the data of the broken
gauge theory, for example those encoding the gauge theoretic structure
of massive W-bosons.

  A generic
point in a patch containing $z=0$ has $s \ne 0$ (and also $t\ne 0$,
but we'll choose to focus on $s$), and the $\bC^*$ action of $\bP^2$
can be used to set $s=1$. On this patch the Weierstrass model becomes
an elliptic fibration over $\bC^2$ with local coordinates $(z,t)$. The
deformed discriminant is a very long expression, but it can be
computed that $z$ does not divide $\Delta_\epsilon$ and that its form
appears to be generic enough to confirm that this is a minimal
Weierstrass model. Furthermore, and a generic $t$
does not lie on the locus $\Delta_\ep=0$. For any such generic $t$ one
can study the Lie algebraic structure of the mentioned elliptic
fibration over the complex $z$-plane.

This geometry, not the blown-up geometry, describes a spontaneously
broken $\mathfrak{g}_2$ gauge theory. We will focus on a particular elliptic
surface with base the $z$-plane, and in this elliptic surface the data of an
$\mathfrak{so}(8)$ symmetry will be evident; outer monodromy induces an
action on this data, which gives rise to gauge algebra $\mathfrak{g}_2$.
Let us first see the $\mathfrak{so}(8)$ structure directly.

We have a family of elliptic fibrations over the $z$-plane
parametrized by the coordinate $t$. To perform a concrete analysis we
fix the complex structure, taking $p_k = t s^{k-1}+l_k s^k+s z^{k-1}+z
t^{k-1}+t^k+z^k$ with $(l_{10},l_{15},l_{18})=(-3,2,1)$ and setting
$s=1$ via the $\bC^*$ action of $\bP^2$. The $l_k$ are chosen such
that the residual discriminant has a $t$ factor,
i.e. $\Delta_r|_{z=0}\sim t$; therefore the $I_0^*$ and $I_1$ loci
have simple normal crossing at $z=t=0$. For any $t$ away from
$\Delta_\epsilon=0$, the Lie algebraic structure
should be evident; we choose $t=i$ for simplicity. Varying $\epsilon$
from $\epsilon=0$ to $\epsilon=.1$ the six marked points at the origin
blossom out, giving a picture
\begin{equation}
\includegraphics{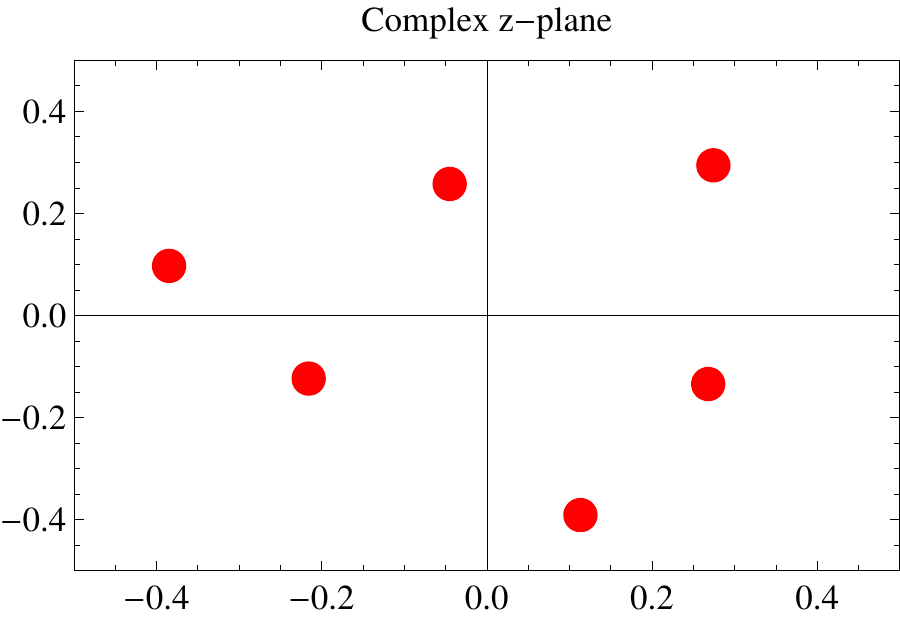}
\end{equation}
at $\ep = .1$. The red dots denote the intersection of $\Delta$ with
the base of this elliptic surface. That is, they are where these
seven-branes intersect the $z$-plane; so we are viewing a cross section
of the seven-branes.

Above each marked point is a singular fiber which has a particular
vanishing one-cycle. In \cite{Grassi:2013kha}, we gave a method for
reading off these vanishing cycles systematically by choosing the
origin to be the base point of the fundamental group and taking
straight line paths of approach from the origin to the marked points,
reading off the vanishing cycles in the process. This can be done
as follows. A Weierstrass equation generically takes the form
$y^2=v_3(x;f,g)$ where $v_3$ is a cubic polynomial in a coordinate $x$
and $f$ and $g$ are appropriate sections which depend on coordinates
in the base. At the base point $z=0$, the roots of $v_3$ are the green
points
\begin{equation}
\begin{tikzpicture}[scale=1]
    \fill[xshift=7cm,thick,color=xdotcolor] (180:10mm) circle (1mm);
    \fill[xshift=7cm,thick,color=xdotcolor] (180-120:10mm) circle (1mm);
    \fill[xshift=7cm,thick,color=xdotcolor] (180+120:10mm) circle (1mm);
    \node at (9.2cm,1.3cm) {$x$};
    \draw[xshift=7cm,thick,->] (180:10mm)+(30:1.3mm) -- +(30:16mm);
    \draw[xshift=7cm,thick,->] (180-120:10mm)+(-90:1.3mm) -- +(-90:16mm);
    \draw[xshift=7cm,thick,->] (180+120:10mm)+(150:1.3mm) -- +(150:16mm);
    \node at (6.6cm,0.7cm) {$\pi_1$};
    \node at (8cm,0cm) {$\pi_2$};
    \node at (6.6cm,-0.7cm) {$\pi_3$};
    \draw[xshift=9cm,thick,yshift=1.0cm] (90:0mm) -- (90:4mm);
    \draw[xshift=9cm,thick,yshift=1.0cm] (0:0mm) -- (0:4mm);
 \end{tikzpicture}
\label{eqn:pixpiypiz}
\end{equation}
which are the ramification points of the elliptic curve described by
the Weierstrass equation, viewed as a two-sheeted cover of the
$x$-plane. The directed paths $\pi_1$, $\pi_2$, and $\pi_3$ describe
one-cycles in the elliptic curve above the base point which satisfy
$\pi_1+\pi_2+\pi_3=0$ in homology. Upon taking a straight path of approach
from the base point to one of the red singular points in the above figure,
$v_3$ changes and two of the three green points in the $x$-plane collide
upon reaching the red singular point in the $z$-plane. This determines a vanishing
one-cycle. Applying this method in this example,
the associated ordered set of vanishing cycles is
\begin{equation}
  Z = \{ \pi_3,\pi_2,\pi_1,\pi_3,\pi_2,\pi_1\}
\end{equation}
beginning with the right-most point in the upper left quadrant and
working clockwise. This is the critical data which serves
as input for a Lie algebraic analysis of the deformed geometry.

Let us briefly review the basic concepts that will be needed below. A
segment in the $z$-plane from one marked point to another, which both
have the same vanishing cycle, is a two-sphere in the total
space\footnote{Here we mean the total space of $X_M$; $M2$-branes
  which wrap those two-cycles become string junctions in the F-theory
  limit.}, since the cycles vanish at the endpoints, giving rise to
the north and south pole. In this elliptic surface this two-sphere has
self intersection $-2$, where there is a contribution of $-1$ from
each endpoint. Picking an appropriate $SL(2,\bZ)$ frame and taking the
type IIb limit, this is just a fundamental string stretched between
two D7-branes.  More generic cycles than this exist, though, which can
end on multiple marked points / seven-branes. In the type IIb
language, these are the string junctions of
\cite{Gaberdiel:1997ud,DeWolfe:1998zf,Mikhailov:1998bx} which end on
$(p,q)$ seven-branes, where $(p,q)$ is the one-cycle vanishing over
the seven-brane in a chosen basis. Of course, in the
M-theory\footnote{Note well that these two cycles in the M-theory
  picture wrapped by $M2$-branes \emph{are not} the two cycles of the
  Coulomb branch.}  picture of the same setup, these are simply
M2-branes wrapping these cycles which have one leg along the fiber and
one leg along the base, giving particles in spacetime.

These more generic junctions also have a generalized intersection
product, which we wrote down in general in \cite{Grassi:2013kha}. It
is simple to see why the intersection product generalizes beyond that
of the string with two endpoints.  Segments of a generalized junction
along which three prongs join are locally pairs of pants in the total
space, and the additional intersections come from the intersections of
the wrapped one-cycles in the elliptic fiber at such ``junction
points.''  Generalized representations come from two manifolds
emanating from the marked points but going off to infinity. Such a
junction $J$ carries an ``asymptotic charge'' $a(J)$, which is just
the one-cycle in the elliptic fiber wrapped by the two-manifold as it
emanates towards infinity; in the type IIb language $a(J)$ is just the
asymptotic $(p,q)$-charge of this state, meaning that it has $p$ units
of of F-string charge and $q$ units of D-string charge. $a(J)$
determines what type of defect/seven-brane the two-manifold could end
on elsewhere in the geometry; if these loci and that extra defect
collide at a codimension two locus, one obtains massless matter
associated to the two-manifold.  A junction $J$ can be represented as
a vector $\in \bZ^N$ where $N$ is the number of defects/seven-branes
in question, but is never the rank of the gauge group. The entries
$J_i$ of $J\in \bZ^N$ are the number of junction prongs ending on the
seven-brane; the formula for the asymptotic charge is simply $a(J) =
\sum_I J_i \, \pi_i$.

Having set the stage, let us proceed to uncover the root lattice of
$\mathfrak{so}(8)$ in this example. States which become massless upon undoing the
deformation by sending $\ep \rightarrow 0$, that is, the W-bosons of
$\mathfrak{so}(8)$, must come from junctions $J$ which shrink to zero size in
that limit. As such, there must be no free end emanating to
infinity; i.e. $a(J)=0$. Furthermore, to reproduce the Cartan data as
expected from comparison to the Coulomb branch, they must have $(J,J)
= -2$ under the intersection product. Concretely, for this ordered set
of vanishing cycles the intersection product and its associated
$I$-matrix can be computed from the
general formula of \cite{Grassi:2013kha}; the $I$-matrix is
\begin{equation}
I\equiv (\cdot,\cdot) = \begin{pmatrix}
-1& 1/2& -1/2& 0& 1/2& -1/2 \\
 1/2& -1& 1/2& -1/2& 0& 1/2 \\
 -1/2& 1/2& -1& 1/2& -1/2& 0 \\
 0& -1/2& 1/2& -1& 1/2& -1/2 \\
 1/2& 0& -1/2& 1/2& -1& 1/2 \\
 -1/2& 1/2& 0& -1/2& 1/2& -1 \\
\end{pmatrix}
\end{equation}
and solving  for the set $R = \{J \in \bZ^6\, | \,a(J) =
\pqvec{0}{0} \,\,\, \text{and} \,\,\, (J,J) = -2 \}$, one obtains
\begin{align}
  R = \{(-1, -1, -1, 0, 0, 0), (-1, -1, 0, 0, 0, -1), (-1, -1, 0, 1,
  1, 0), (-1, 0, 0, 0, -1, -1), (-1, 0, 0, 1, 0, 0), \nonumber \\ (-1, 0, 1, 1, 0,
  -1), (0, -1, -1, -1, 0, 0), (0, -1, -1, 0, 1, 1), (0, -1, 0, 0, 1,
  0), (0, 0, -1, -1, -1, 0), \nonumber \\ (0, 0, -1, 0, 0, 1), (0, 0, 0, -1, -1,
  -1), (0, 0, 0, 1, 1, 1), (0, 0, 1, 0, 0, -1), (0, 0, 1, 1, 1, 0),\nonumber \\
  (0, 1, 0, 0, -1, 0), (0, 1, 1, 0, -1, -1), (0, 1, 1, 1, 0, 0), (1,
  0, -1, -1, 0, 1), (1, 0, 0, -1, 0, 0),\nonumber \\ (1, 0, 0, 0, 1, 1), (1, 1, 0,
  -1, -1, 0), (1, 1, 0, 0, 0, 1), (1, 1, 1, 0, 0, 0)\}
\end{align}
where a particular entry of a vector indicates the number of prongs
on the respective marked point, according to the ordering of $Z$. Note
that the set has $|R|=24$, matching precisely the number of roots of $\mathfrak{so}(8)$.
The subset
\begin{equation}
SR = \{[0, 0, 0, 1, 1, 1],
 [0, 0, 1, 0, 0, -1],
 [0, 1, 0, 0, -1, 0],
 [1, 0, -1, -1, 0, 1]\}
\end{equation}
is a set of four simple root junctions, as verified by checking that
the $I$-matrix of these junctions is precisely negative of
the Cartan matrix for $\mathfrak{so}(8)$
\begin{equation}
-A_{ij} = (J_i,J_j) = \begin{pmatrix}
-2 & 1 & 0 & 0 \\
1 & -2 & 1 & 1 \\
0 & 1 & -2 & 0 \\
0 & 1 & 0 & -2
\end{pmatrix} \qquad \text{for} \qquad J_i \in SR.
\end{equation}
One can also check, given the full set of roots above, that there is a
highest root determined by these simple roots, and that the associated
level diagram generated by Freudenthal's recursion formula\footnote{Given
 the highest weight of a Lie algebra representation and the simple roots,
Freudenthal's recursion formula can be utilized to generate all of
the weights in a representation, along with their multiplicities. See
\cite{Grassi:2013kha} for further discussion.} is of the
correct topology. Accordingly, the levels of the roots are
\begin{center}
\scalebox{.8}{\begin{tabular}{c|c|c|c}
Level & $J$ & $(J,J)$ & Multiplicity \\ \hline
$0$ & $(1, 1, 1, 0, 0, 0)$ & $-2$ & $1$ \\
$1$ & $(1, 1, 0, 0, 0, 1)$ & $-2$ & $1$ \\
$2$ & $(1, 1, 0, -1, -1, 0)$ & $-2$ & $1$ \\
$2$ & $(1, 0, 0, 0, 1, 1)$ & $-2$ & $1$ \\
$2$ & $(0, 1, 1, 1, 0, 0)$ & $-2$ & $1$ \\
$3$ & $(1, 0, 0, -1, 0, 0)$ & $-2$ & $1$ \\
$3$ & $(0, 1, 1, 0, -1, -1)$ & $-2$ & $1$ \\
$3$ & $(0, 0, 1, 1, 1, 0)$ & $-2$ & $1$ \\
$4$ & $(1, 0, -1, -1, 0, 1)$ & $-2$ & $1$ \\
$4$ & $(0, 0, 1, 0, 0, -1)$ & $-2$ & $1$ \\
$4$ & $(0, 1, 0, 0, -1, 0)$ & $-2$ & $1$ \\
$4$ & $(0, 0, 0, 1, 1, 1)$ & $-2$ & $1$ \\
$5$ & $(0, 0, 0, 0, 0, 0)$ & $0$ & $4$ \\
$6$ & $(0, 0, 0, -1, -1, -1)$ & $-2$ & $1$ \\
$6$ & $(0, 0, -1, 0, 0, 1)$ & $-2$ & $1$ \\
$6$ & $(0, -1, 0, 0, 1, 0)$ & $-2$ & $1$ \\
$6$ & $(-1, 0, 1, 1, 0, -1)$ & $-2$ & $1$ \\
$7$ & $(0, 0, -1, -1, -1, 0)$ & $-2$ & $1$ \\
$7$ & $(0, -1, -1, 0, 1, 1)$ & $-2$ & $1$ \\
$7$ & $(-1, 0, 0, 1, 0, 0)$ & $-2$ & $1$ \\
$8$ & $(0, -1, -1, -1, 0, 0)$ & $-2$ & $1$ \\
$8$ & $(-1, 0, 0, 0, -1, -1)$ & $-2$ & $1$ \\
$8$ & $(-1, -1, 0, 1, 1, 0)$ & $-2$ & $1$ \\
$9$ & $(-1, -1, 0, 0, 0, -1)$ & $-2$ & $1$ \\
$10$ & $(-1, -1, -1, 0, 0, 0)$ & $-2$ & $1$ \\
\end{tabular}}
\end{center}
This includes, for example, a symmetry on the level diagram indicative
of $\mathfrak{so}(8)$ triality; notice that there exist a number of
levels with three roots. $M2$-branes on these two-manifolds, which are
string junctions in F-theory, give rise to particle states filling out
an adjoint of $\mathfrak{so}(8)$. However, we know that this geometry
exhibits instead a broken $\mathfrak{g}_2$ gauge symmetry due to outer
monodromy; in \cite{Grassi:2013kha} we showed that codimension two
singularities can, in fact, induce an action on junctions upon moving
in a family of elliptic surfaces, and that the associated algebraic
automorphism is a $\bZ_3$ outer automorphism that turns
$\mathfrak{so}(8)$ into $\mathfrak{g}_2$. In fact, the local geometry 
studied in \cite{Grassi:2013kha} is just the geometry we have considered
(with the specific $p_k$) near the locus $z=t=0$, $s=1$.

Finally, though we have a broken $\mathfrak{g}_2$ gauge theory, let us study
a few non-adjoint representations of $\mathfrak{so}(8)$ at the level
of junctions for the sake of illustration. These arise from
two-manifolds $J$ in this elliptic surface which carry some asymptotic
charge $a(J)$. For simple representations, all $J$ making up the
representations will have the same self-intersection $(J,J)$; this is
not true, of course, for higher dimensional representations, whose
weights may not all have the same length.  Studying three different
sets of junctions, each with\footnote{Note that in the code referenced
  in appendix \ref{sec:code} we choose a basis of one-cycles such that
  $\pi_1=\pqvec{1}{0}$ and $\pi_3 = \pqvec{0}{1}$ in order to perform
  concrete computations; from this one computes $\pi_2=\pqvec{-1}{-1}$
  from the fact that $\pi_1+\pi_2+\pi_3$ is trivial in homology.}
$(J,J)=-1$ but with $a(J) = \pi_1$, $a(J)=\pi_2$ and $a(J)=\pi_3$,
gives three different representations.  From these sets the level
diagrams of the junctions can be determined.  They are
\begin{center}
  \centering
\scalebox{.65}{
  \begin{tabular}{c|c|c|c|c|c|c|c|}
    $a(J)$ & Level 0 &Level 1 &Level 2 &Level 3 &Level 4 &Level 5 &Level 6  \\ \hline
    $\pi_1$ &$(0, 1, 1, 0, -1, 0)$ & $(0, 0, 1, 0, 0, 0)$ &$(0, 0, 0, 0, 0, 1)$ & $(0, 0, 0, -1, -1, 0)$, $(-1, 0, 1, 1, 0, 0)$ & $(-1, 0, 1, 0, -1, -1)$ & $(-1, 0, 0, 0, -1, 0)$ & $(-1, -1, 0, 0, 0, 0)$\\
    $\pi_2$ & $(0, 0, 1, 1, 0, 0)$ & $(0, 0, 1, 0, -1, -1)$ & $(0, 0, 0, 0, -1, 0)$ & $(0, -1, 0, 0, 0, 0)$, $(1, 0, 1, 1, -1, -1)$ & $(-1, -1, 1, 1, 0, -1)$ & $(-1, -1, 0, 1, 0, 0)$ & $(-1, -1, 0, 0, -1, -1)$ \\
    $\pi_3$ & $(1, 0, 0, 0, 0, 0)$& $(0, 0, 1, 1, 0, -1)$& $(0, 0, 0, 1, 0, 0)$& $(0, 0, 0, 0, -1, -1)$,$(0, -1, 0, 1, 1, 0)$&$(0, -1, 0, 0, 0, -1)$ &$(0, -1, -1, 0, 0, 0)$& $(-1, -1, 0, 1, 0, -1)$\\ \hline
  \end{tabular}}
\end{center}
where the highest weight junctions are those at level $0$. These
junctions are vectors in $\bZ^6$, but since they represent
some representation of the $D_4$ algebra, which is of rank $4$,
there must be a linear map from $\bZ^6$ to $\bZ^4$ which maps
junctions to weights in the Dynkin basis. We wrote down such
a general map in \cite{Grassi:2013kha}, which in this example is
computed to be
\begin{equation}
  F=\begin{pmatrix}
    0 & 0 & 0 & 1 & 0 & 1 \\
    0 & 0 & 1 & -1 & 1 & -1 \\
    0 & 1 & -1 & 1 & -1 & 0 \\
    1 & -1 & 0 & 0 & -1 & 1
  \end{pmatrix}.
\end{equation}
Applying this map to the highest weight junctions above, these eight
dimensional representations have Dynkin labels given by $(0,0,1,0)$,
$(1,0,0,0)$, and $(0,0,0,1)$. Recall that $\mathfrak{so}(8)$ has three eight
dimensional representations, which are permuted by triality. Their
highest weights have precisely these Dynkin labels; these sets of
junctions fill out those representations.

\clearpage
\section{Gauge Theories and Singularity Deformation}
\label{sec:deformations}

We have just seen how singularity deformation describes the spontaneous
breaking of gauge symmetry in F-theory and how structures of the broken
gauge theory can be identified in the deformed geometry.

In this section we would like to make some brief, but we think
important, comments about physics, the relationship between the
geometric and gauge theory moduli spaces, and also the only known
description for describing both massless and massive gauge states in
F-theory.  We will then develop a general framework for performing
deformations which allow one to see these structures in the geometry.

\subsection{The F-theoretic and M-theoretic Physics of Deformation}
\label{sec:deformation physics}
For the sake of clarity we would like to again state the points about
the M-theoretic and F-theoretic physics of singularity deformation;
some may exist elsewhere in the literature. Though we will utilize
$d=3,4$ language exclusively, many of these comments apply to other
dimensions as well.

\vspace{.5cm}
\noindent \emph{Gauge Theoretic and Geometric Moduli Spaces}

A $d=4$ compactification of F-theory on $X$ and its defining M-theory
compactification on an elliptically fibered Calabi-Yau variety $X_M$
give rise to $d=4$ $\cN=1$ and $d=3$ $\cN=2$ gauge theories,
respectively, and movement in these moduli spaces occurs by movement
in corresponding geometric moduli spaces.

Since the resolution of singular fibers in $X_M$ is not consistent
with the limit of vanishing fiber $X_M\rightarrow X$ which defines
F-theory, this branch of the moduli space does not exist\footnote{A
  more physical way to state its non-existence is as follows. In
  F-theory compactifications gravitons can propagate in $\bR^{3,1}$
  and also the six-dimensional base $B$ of $X$; if the rational curves
  of resolution were to appear gravitons could propagate in them,
  also, for a total of twelve dimensions of graviton
  propagation. Since no such twelve-dimensional theory exists, it must
  be impossible to do this in F-theory. In M-theory on $\bR^{2,1}
  \times X_M$ the smooth elliptic fibers and also the rational curves
  of resolution are available for graviton propagation, for a total of
  $11$ dimensions, which is of course perfectly consistent.} in F-theory, and thus
\begin{align}
  \text{\emph{1) The resolution of singular fibers does not describe
      the spontaneous }}\nonumber \\ \text{ \emph{breaking of any
      gauge symmetry in F-theory}}. \nonumber
\end{align}
\noindent Since it can break the
$d=3$ $\cN=2$ gauge theories but not the $d=4$ $\cN=1$ gauge theories,
this is sufficient to show that those K\" ahler moduli which perform
this breaking must correspond to the scalars obtained from the
dimensional reduction of the $4^{th}$ component of the gauge field, as
is well known.

Alternatively, both $X_M$ and $X$ can undergo deformation of complex
structure. This means that the associated moduli can break both
the $d=3$ $\cN=2$ gauge theories of M-theory on $X_M$ and also the
$d=4$ $\cN=1$ gauge theories of F-theory on $X$. As such, these
\begin{align}
  \text{\emph{2) Complex structure moduli determine the expectation values
    of scalar fields in}}  \nonumber \\
  \text{\emph{$d=4$ chiral multiplets and their dimensional reductions to three dimensions,}} \nonumber
\end{align}
\noindent
though it remains to work out a general map.  Thus, it is complex
structure deformations that determine the spontaneous breaking of
gauge theories arising in four-dimensional F-theory compactifications.

\clearpage
\vspace{.5cm}
\noindent \emph{Gauge States in F-theory are String Junctions}

An immediately corollary of the fact that resolution of singular
fibers does not break any gauge symmetry in F-theory is that
\begin{equation}
  \text{\emph{3) Resolution of singular fibers cannot describe massive gauge states in F-theory,}}
\end{equation}
\noindent for example
massive W-bosons. This is not ideal, as a preferred description of the
theory should be able to describe both massive and massless gauge
states!

Alternatively, as complex structure deformation does describe the
spontaneous breaking of gauge symmetry in F-theory, it does describe
those massive W-bosons (and their superpartners). In the deformed
M-theory picture, these are $M2$-brane states wrapped on two-cycles
with one leg on the fiber and one leg on the base; in the F-theory
limit these objects lose a dimension and become string
junctions. Thus, from these arguments we see unambiguously that
\begin{equation}
  \text{\emph{4) Gauge states in F-theory arising from
      singular fibers are string junctions, possibly in a massless limit.}}
\end{equation}
It is reasonable to expect that significant progress in F-theory can
be made by utilizing this description, since it exists both in
a defining M-theory compactification and its F-theory limit.

In addition, there are known examples
  \cite{Grassi:2011hq} of six-dimensional F-theory compactifications
  where anomaly cancellation requires a multiplicity
  factor for matter fields which is not captured by a simple resolution, since
  this factor is related to the versal deformation space near the
  singularities.  It would be interesting to see if our techniques
  correctly capture this multiplicity. Furthermore, the Higgs
  couplings in the low energy effective action of a four-dimensional
  $\cN=1$ theory are critical for a proper understanding of the
  physics; in F-theory the expectation values of Higgs fields, which affect
  the Higgs couplings,
  typically depend on complex structure moduli, and never on the K\"
  ahler moduli of resolution. 

\vspace{.5cm}
\noindent \emph{Higgs and Coulomb Branches of
  Singularity Deformation}

To what subgroups can singularity deformation break $G$ in F-theory?
This depends as usual on the representation which performs the
Higgsing and also which of its components receive expectation values,
but the possibility exists for both Higgs branches and Coulomb
branches in $d=4$ gauge theories. In particular, if the field which
obtains an expectation value is the scalar component of a
four-dimensional adjoint chiral multiplet, the theory is broken to
$U(1)^{rk(G)}$; this is a four-dimensional Coulomb branch which can
also exist in three dimensions, where one might call it the ``M-theory
Coulomb branch of singularity deformation,'' in contrast to its
resolution counterpart. Thus,
\begin{align}
  \text{\emph{5) There can be an M-theory Coulomb branch which lifts
      to F-theory;}}\nonumber \\ \text{\emph{ It is the one of
      deformation, not resolution.}} \nonumber
\end{align}
On this branch non-abelian gauge states are described by string
junctions. Since the Cartan $U(1)$'s of $G$ have not been broken, this
gauge theoretic description geometrically requires the existence of
deformations which do not change $h^{1,1}(X)$ or $h^{1,1}(X_M)$; these
are known to exist in certain examples. For example, in the threefold
case see \cite{Katz:1996ht}; there these deformations completely break
the non-abelian part of the gauge theory, but a $U(1)^{rk(G)}$
subgroup remains, along with a finite collection of conifolds.

Alternatively, depending on the representation theory there can also
be Higgs branches in F-theory compactifications and their defining
M-theory compactifications where the gauge symmetry is completely
Higgsed to nothing. Whether a particular deformation gives a Higgs
branch or a Coulomb branch depends on the details of the geometry.

\subsection{Deformation Types for Weierstrass Models}
\label{sec:weierstrass def}
Having emphasized important aspects of the F-theoretic and M-theoretic
physics of deformation, let us discuss certain deformation types of
Weierstrass models which will be useful. There has been much study in
the mathematics literature of resolution and deformations of the
surface A-D-E (Kleinian) singularities
\cite{Arnold,Slodowy,Rossi,KatzMorrison} and also of complete
intersections isolated points \cite{Looijenga}. The 
deformations of relevance in this context  are the deformations of the elliptic fibration $X \to B$
around the singular loci.

Consider a Weierstrass model for a Calabi-Yau elliptic fibration
$\pi:X\rightarrow B$.
The defining equation for such a variety is given by
\begin{equation}
  \label{eq:Weierstrass}
  y^2 = x^3 + f\, x + g
\end{equation}
with sections $f \in \Gamma(K_B^{-4})$ and $g \in
\Gamma(K_B^{-6})$. The singular fibers occur over points in the base
in the discriminant locus $\Delta \equiv 4f^3 + 27g^2 = 0$.  For the
singularities relevant for F-theory compactifications with non-abelian
gauge symmetry, the discriminant
takes the form
\begin{equation}
  \label{eq:singulardiscriminant}
  \Delta = z^N \, \Delta_r
\end{equation}
where $\Delta_r$ is easily computed in examples. The singular fiber
above the locus $z=0$ is determined by Kodaira's classification of
singular fibers in codimension one. In many cases this singular fiber
corresponds to an ADE singularity\footnote{An interesting fact is that
  $N$ is never the rank of the gauge group; as such root junctions are
  represented as vectors in $\bZ^N$ which nevertheless only span a
  $rk(G)$-dimensional subspace.}; the only counterexample is a type II
fiber, which we will explicitly show does not carry a gauge algebra
using deformation.  We call $\Delta_r$ the \emph{residual
  discriminant}, which itself may be a reducible subvariety in
$B$. Its form determines the precise structure of singular fibers in
codimension two and three, as well as the associated physics. For
example, though the representation theory of matter fields is
determined by codimension one data, it is the structure of $\Delta_r$
that determines which matter fields become massless in codimension
two. See \cite{Grassi:2013kha} for discussion on this
point.

For a Weierstrass model\footnote{Similar deformations could also be
  done for the sections $a_n$ of a Tate model.}, it is simple to see
an organizing principle for the deformations: one can deform both $f$
and $g$ as
\begin{equation}
  f \mapsto f + \ep_f, \qquad g \mapsto g + \ep_g
\end{equation}
where $\ep_f$ and $\ep_g$ are sections of the same bundles as
$f$ and $g$, of course.  Such a deformation deforms the discriminant
locus as
\begin{equation}
\label{eqn:deformeddiscriminant}
\Delta \mapsto \Delta + 4 \epsilon _f \left(3 f^2+3 f \epsilon _f+\epsilon _f^2\right) + 27 \epsilon _g \left(\epsilon _g+2 g\right)
\end{equation}
We call such a deformation a \emph{$fg$-deformation} and will call the
deformation of the discriminant the \emph{deformed discriminant}. In
some cases these general deformations will not be needed to uncover
gauge structure and we will consider an \emph{$f$-deformation} or a
\emph{$g$-deformation}, defined by $\ep_g=0$ and $\ep_f=0$,
respectively; that is, these deform either $f$ or $g$, but not
both. The deformed discriminant is easily derived from the more
general form.

 Having discussed the general ways to deform a Weierstrass
equation --- by deforming $f$, $g$, or both --- we now categorize
certain types of deformations to set a common language for discussing
examples.  Taking a singular variety with $\Delta = z^N \, \Delta_r$,
a general deformation gives $\Delta \mapsto z^N \, \Delta_r +
\Delta_\ep$ for some $\Delta_\ep$.

Suppose that $z$ does not divide $\Delta_\ep$. Then in small neighborhoods
of $z=0$ away from $\Delta_\ep \ne 0$ and $\Delta_r \ne 0$ the deformed
discriminant is
\begin{equation}
z^N \, \Delta_r + \Delta_\ep = \Delta_r \, (z^N + \Delta_\ep / \Delta_r)
\end{equation}
and we see that $\Delta \sim (z^N + \tilde \ep)$ for $\tilde \ep
\equiv \Delta_\ep/\Delta_r$. This gives a family of elliptic surfaces
with base the complex $z$-plane, where the family is determined by
base coordinates via $\tilde \ep$; for a generic member of the family,
$\tilde \ep\in \bC^*$.  We call such a deformation a \emph{completely
  Higgsing deformation} since for a generic member of the family the
singularity at $z=0$ has been completely smoothed out; the associated
non-abelian part of the gauge theory has been Higgsed\footnote{Though
  in certain examples it may be possible for abelian subgroups to survive.}. More
specifically, the $N$ coincident components of $z^N$ have been
deformed into $N$ distinct components at the $N^\text{th}$ roots of
$\tilde \ep$.  The singular fiber above each component has a vanishing
one-cycle, and by identifying an ordered set of such vanishing cycles,
the new finite volume two-cycles whose structure reproduces the
structure of the Higgsed gauge algebra can be systematically
constructed \cite{Grassi:2013kha}. We call $f$-deformations and
$g$-deformations which are completely Higgsing \emph{completely
  Higgsing $f$-deformations} and \emph{completely Higgsing
  $g$-deformations}, respectively. Finally, we note that a completely
Higgsing deformation may not actually smooth all singularities: we
have named it to denote the complete Higgsing of a particular
codimension one gauge theory along $z=0$; there may, of course, be
other gauge theories on other components of the discriminant.

In some cases it will be useful to seek deformations which keep
certain structures intact. We will not use such deformations in this
work, but one was utilized in the $\mathfrak{g_2}$ example of
\cite{Grassi:2013kha} and will be utilized in our upcoming works
\cite{GrassiHalversonShaneson:Math,GrassiHalversonShaneson:PhysicsCodim2}.
We call an $fg$-deformation with

\begin{equation}
  f \mapsto f + \ep_{f,r} \, \Delta_r \qquad \qquad \text{and} \qquad \qquad g \mapsto g + \ep_{g,r} \, \Delta_r
\end{equation}

a \emph{$\Delta_r$-deformation}. In such a case the deformed discriminant is
\begin{equation}
\Delta = \Delta_r \left[z^N + 4\epsilon _{f,r} \left(3 f \Delta _r \epsilon _{f,r}+\Delta _r^2
   \epsilon _{f,r}^2+3 f^2\right) + 27 \epsilon _{g,r} \left(\Delta _r \epsilon _{g,r}+2 g\right)\right]
\end{equation}
and we see the additional property enjoyed by a
$\Delta_r$-deformation: after deformation, $\Delta_r$ is still a
component of the discriminant. Of course, in a compact model such a
deformation requires the existence of appropriate sections
 $\ep_{f,r}\in \Gamma(\cO(N\,D_z) \otimes K_B^8)$ and
$\ep_{g,r} \in \Gamma(\cO(N\,D_z) \otimes K_B^6)$, where $D_z$ is the
divisor class of the $z=0$ locus.  If $\Delta_r$ has a particular
component $\Delta_i$ whose codimension two intersection with $z=0$ is
of interest, we may also consider a \emph{$\Delta_i$-deformation},
defined by $\ep_f = \ep_{f,i} \, \Delta_i$ and $\ep_g = \ep_{g,i} \,
\Delta_i$, the existence of which requires the existence of
appropriate sections $\ep_{f,i}$ and $\ep_{g,i}$. In such a case
$\Delta_i$ remains a component of the discriminant after
deformation. In some cases it may be that a $\Delta_i$-deformation
exists, even if a $\Delta_r$-deformation does not. Note that both of
these cases can be restricted to $f$-deformations and $g$-deformations
by setting $\ep_{g,i}$ and $\ep_{f,i}$ to zero, respectively.

We will find that $\Delta_r$- and $\Delta_i$-deformations are useful
for studying outer automorphisms on codimension one algebras induced
by monodromy around codimension two singularities. As in our previous
work, we will call such monodromy \emph{O-monodromy}, which we
emphasize is different than the monodromy induced by the
Picard-Lefschetz action on vanishing cycles. The general method we
will employ to study O-monodromy is to deform a codimension one
singularity and analyze its Lie algebraic structure, and then to
encircle another codimension one locus which intersects $z=0$ at the
codimension two point of interest prior to deformation. Herein lies
the advantage of $\Delta_r$- and $\Delta_i$-deformations: after
deformation, $\Delta_r$ or $\Delta_i$ are still components of the
discriminant, and one of these is the codimension one locus which will
be encircled to study O-monodromy. If we had performed a generic
deformation which was not a $\Delta_r$ or $\Delta_i$ deformation, it
would not be clear what to encircle.

\subsection{Local vs. Global Deformations and non-Higgsable Clusters}\label{NHC}
\label{sec:global to local}
While in many examples we use the above technology to deform local
Weierstrass models, our techniques can be used also in compact
geometries, in particular for global compactifications of M-theory and
F-theory. In fact, we have already presented such a compact
example in section \ref{sec:glimpse}.

In the following subsection we discuss two examples demonstrating
interesting features of global and local deformations. In the first,
the local deformation is actually the restriction of a global
deformation. In the second example we show how to use our local
techniques, together with a global geometry analysis, to deduce global
information about gauge algebras and matter representations, even in
cases where no global deformation is possible. These are
``non-Higgsable clusters" defined and classified in
\cite{Morrison:2012np}.

\vspace{.5cm}
\noindent \emph{A Simple Example of a Global to Local Map} \\
\noindent Consider an F-theory compactification to four dimensions on
an elliptically fibered Calabi-Yau fourfold with $B=\bP^3$. We
consider a Weierstrass model; since $K_B=\cO_B(-4)$ we have $f \in
\Gamma(\cO_B(16))$ and $g \in \Gamma(\cO_B(24))$ in homogeneous
coordinates $(z,x_1,x_2,x_3)$ on $\bP^3$. Suppose that $f$ and $g$ are
chosen such that they have order of vanishing zero around $z=0$, but
the discriminant has order of vanishing $5$. In such a case there is
an $I_5$ fiber above a generic point in the locus $z=0$, and the
discriminant takes the form
\begin{equation}
\Delta = z^5 \, \Delta_r.
\end{equation}
This is appropriate for a global F-theory compactification with an
$\mathfrak{su}(5)$ factor in the gauge algebra.

We would like to perform a complex structure deformation of the
compact manifold which is a completely Higgsing
$g$-deformation. Specifically, we consider the deformation given by
\begin{equation}
\wgdef{\ep\, [x_1^{24} + x_2^{24}+x_3^{24}]} \qquad \qquad \ep \in \bC^*,
\end{equation}
which is an allowed one-parameter deformation of the global
geometry. The deformed discriminant is
\begin{equation}
  \Delta = z^5\, \Delta_r + 54 \, \ep\, g \left(x_1^{24}+x_2^{24}+x_3^{24}\right) +27\,\epsilon^2\,
  \left(x_1^{24}+x_2^{24}+x_3^{24}\right)^2
\end{equation}
The structure of this equation matches the general form
(\ref{eqn:deformeddiscriminant}), $\Delta$ is an $I_1$ locus, and the
non-abelian part of the gauge theory along $z=0$ is completely
Higgsed.

How does this deformation map to a deformation of a local Weierstrass
model, as will be studied throughout this work? Since
$z=x_1=x_2=x_3=0$ is not in $\bP^3$ any patch containing $z=0$ must
have at least one $x_i\ne 0$. Suppose it is $x_1\ne 0$. On that patch,
the $\bC^*$ action of $\bP^3$ can be used to set that $x_1=1$,
and at every point on that patch the deformation appears as
\begin{equation}
  \wgdef{\tilde \ep}
\end{equation}
with $\tilde \ep=\ep \, [1+x_2^{24}+x_3^{24}]$; this gives a family of
elliptic surfaces over the $z$-plane determined by $x_2,x_3$.  For
generic $x_2,x_3$, $(1+x_2^{24}+x_3^{24})\ne 0$; this means that a
generic member of the family of elliptic surface is deformed by this
global deformation. For a particular member of the family, $\tilde
\ep$ is simply a number and we have obtained a local description of
the elliptic fibration over the $z$-plane at this point.

In many examples we will simply deform $g$ by adding a number $\ep \in
\bC^*$; here we see have seen how deformations of such a local model
are realized in a global context.

\vspace{.7cm}
\noindent \emph{From local deformation to global information, even when a global
deformation does not exist: \\ the case of non-Higgsable clusters}

\vspace{.3cm}
\noindent In \cite{Morrison:2012np} the authors consider Calabi-Yau
varieties $X \to B$ which are resolutions of maximally Higgsed
(minimal) Weierstrass models. They analyze all connected
configurations of curves $C$ of negative self-intersection in $B$
which carry non-Higgsable gauge algebra: these curves are necessarily
smooth, rational and of negative self intersection, say $-m$. Since
the Weierstrass model is assumed to be minimal, $m \leq 12$.  For
example, if $B= \bF_m$ is a Hirzebruch surface with $m \geq 3$, then
the ``infinity section" $C_\infty$ with ${C_\infty}^2=-m$ must carry a
gauge algebra, and are necessarily non-Higgsable clusters.

Under general conditions, assumed in \cite{Morrison:2012np}, the
vanishing orders of $f$ and $g$ near the curve $C$ are also determined
by the geometry and can be calculated explicitly.  Table 2 in
\cite{Morrison:2012np} describes the (connected) non-Higgsable
clusters consisting of a single curve:

\begin{center}
\begin{tabular}{c|c|c|c}
$-m$ & Gauge Algebra & Matter & $(f, g, \Delta)$ \\ \hline
$ -3 $ & $ \mathfrak{su}(3) $ & $ 0$ & $ (2,2,4) $ \\
$ -4 $ & $ \mathfrak{so}(8) $ & $ 0 $ & $ (2,3,6) $ \\
$ -5 $ & $ \mathfrak{f}_4 $ & $ 0 $ & $ (3,4,8) $ \\
$ -6 $ & $ \mathfrak{e}_6 $ & $ 0 $ & $ (3,4,8) $ \\
$ -7 $ & $ \mathfrak{e}_7 $ & $ \frac{1}{2}{\mathbf 56} $ & $ (3,5,9) $ \\
$ -8 $ & $ \mathfrak{e}_7 $ & $ 0 $ & $ (3,5,9) $ \\
$ -12 $ & $ \mathfrak{e}_8 $ & $ 0 $ & $ (4,5,10) $ \\
\end{tabular}
\end{center}

We show that our method can be applied also in this situation, namely
even when a global deformation (Higgsing) does not exist.  Let
$\mathcal{V}$ in $B$ be a sufficiently small neighborhood (in the
complex topology) such that $C \subset \mathcal{V}$. Let $\mathcal{D}
\subset \mathcal{V}$ a divisor which intersects $C$ in a smooth point
and let $ \mathcal{U}= \mathcal{V} \setminus \mathcal{D} $. Then the
elliptic fibration restricted to $ \mathcal{U}$ is an affine
Weierstrass model $W_0 \to \mathcal{U}$: $y^2=x^3+f(z)\,x+g(z)$, where
$z$ is the local parameter of $C$, and the vanishing orders of $f$ and
$g$ along $C$ are described in the fourth column. We can then deform
this affine Weierstrass model as described in the previous paper
\cite{Grassi:2013kha}; see section \ref{sec:su3} for the case
$m=3$. Note that when $m \neq 5, 7$ the curve $C$ does not intersect
the other components of the discriminant: we then obtain, in
particular, the gauge algebras in the second column of the above Table.
If $m=5$, the same geometric analysis described in
\cite{Morrison:2012np} tells us the points are branched points of an
outer monodromy and that the contribution to the matter along $C$ is
zero. In the forthcoming paper
\cite{GrassiHalversonShaneson:PhysicsCodim2} we derive the explicit
matter contribution if $m=7$, as well as the non-localized matter for
$m=5$.

\clearpage
\section{Codimension One Gauge Structure: Less Studied Fibers}
\label{sec:less studied fibers}
The codimension one fibers utilized in gauge theories or F-theory are
typically of type $I_n$, $I_n^*$, $IV^*$, $III^*$, $II^*$,
corresponding to gauge algebras $\mathfrak{su}(N)$, $\mathfrak{so}(2(N+4))$, $\mathfrak{e}_6$, $\mathfrak{e}_7$,
and $\mathfrak{e}_8$, respectively. For $I_n$, $I_n^*$, $IV^*$, $III^*$, and
$II^*$ fibers the relevant data has already been obtained in the
literature; for explicit deformations of Calabi-Yau elliptic
fibrations, see \cite{Grassi:2013kha}, and for other examples and
ordered sets of vanishing cycles, see some of the original literature
on string junctions \cite{Gaberdiel:1997ud,DeWolfe:1998zf}.

In this section we will deform Kodaira's other codimension one
singular fibers\footnote{Links
to computational codes and illustrative films for these examples
are provided in appendix \ref{sec:code}.}, of type $IV$, $III$, and $II$ and recover the
corresponding gauge algebras $\mathfrak{su}(3)$, $\mathfrak{su}(2)$ and $\emptyset$,
respectively. Interestingly, these $\mathfrak{su}(3)$ and $\mathfrak{su}(2)$ gauge algebras
arise from four and three sets of seven-branes, respectively, as
opposed to the three and two sets in type IIb; this is an
intrinsically F-theoretic realization of these algebras.

\subsection{SU(3) Gauge Symmetry from a Type IV Fiber}
\label{sec:su3}
In this example we study the gauge symmetry 
realized by a type IV fiber in codimension one. We find this example
intriguing: by deforming the geometry we will see the appearance of an
$\mathfrak{su}(3)$ gauge algebra realized by string junctions ending
on \emph{four} seven-branes, as opposed to the standard three
seven-branes in the case of an $I_3$ fiber, which becomes three
coincident D7-branes in the weakly coupled type IIb picture.  In the case of threefolds or higher dimension,  a type IV fiber can either realize $\mathfrak{su}(3)$ or
    $\mathfrak{sp}(1)\cong \mathfrak{su}(2)$ due to outer monodromy.

We start with the case of a surface, or equivalently with the analysis of a singular fiber of type IV around a general point of the discriminant locus. The general Weierstrass form, the singular fiber can be realized by
\begin{equation}
  \label{eq:IVcodim1}
  f = z^2 \qquad g = z^2.
\end{equation}
and the discriminant is $\Delta = z^4 \left(4 z^2+27\right)$.
Since we performed a completely Higgsing $g$-deformation in section
\ref{sec:glimpse}, let us utilize a completely Higgsing $fg$-deformation
for the sake of illustration. The $fg$-deformation $\wdef{\ep}{\ep}$ has
an associated deformed discriminant
\begin{equation}
  \Delta = \left(z^2+\epsilon \right)^2 \left(4 z^2+4 \epsilon +27\right).
\end{equation}
We see that this deformation is not completely Higgsing. Instead, we
study the deformation
\begin{equation}
  \label{eq:IVcodim1def}
  \wdef{2 \ep}{\ep},
\end{equation}
with associated deformed discriminant
\begin{equation}
  \Delta = 4 z^6+27 z^4 + 6 \, \ep \, z^2 \left(4 z^2+9\right)  + 3 \, \ep^2\, \left(16 z^2+9\right)  + 32\, \ep^3
\end{equation}
which appears to have broken the degeneracy. In all, the single
discriminant component with multiplicity four has been deformed into
four discriminant components with multiplicity one.

Taking $\ep = .1$ and going to a neighborhood of $z=0$ in order to see
the four deformed components (and not the other two $z$-roots of
$\Delta$) the intersection of the deformed discriminant with the
$z$-plane is given by
\begin{center}
\includegraphics[scale=.8]{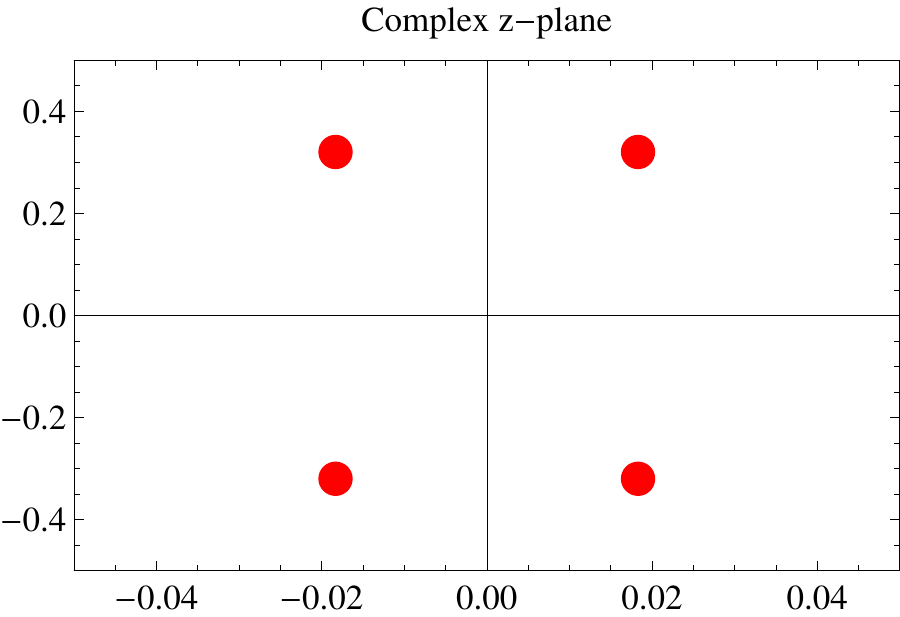}
\end{center}
and we see that, indeed, the gauge theory previously at $z=0$ has been
completely Higgsed.  Picking the base point to be to at the origin and
following straight line paths of approach to the singular fibers,
beginning with the upper left and working clockwise, determines an
ordered set of vanishing cycles
\begin{equation}
  Z_{IV} = \{\pi_1,\pi_3,\pi_1,\pi_3\}.
\end{equation}
>From the generic formula of \cite{Grassi:2013kha}, the
$I$-matrix for this ordered set of vanishing cycles is
\begin{equation}
I = (\cdot, \cdot) = \begin{pmatrix}
-1 & 1/2 & 0 & 1/2 \\
1/2 & -1 & -1/2 & 0 \\
0 & -1/2 & -1 & 1/2 \\
1/2 & 0 & 1/2 & -1
\end{pmatrix}.
\end{equation}
Per usual, the roots are $\{ J \in \bZ^4 \,\, | \,\, (J,J) = -2 \,\,\,
\text{and}\,\,\, a(J) = \pqvec{0}{0}\}$. They are given by
\begin{equation}
(-1, -1, 1, 1), (-1, 0, 1, 0), (0, -1, 0, 1), (0, 1, 0, -1), (1, 0, -1, 0), (1, 1, -1, -1), (0, 0, 0, 0), (0, 0, 0, 0)
\end{equation}
to which the Cartan generators $ (0, 0, 0, 0), (0, 0, 0, 0)$ have been
added. For one possible set of simple roots, the root diagram is given
by
\begin{center}
\begin{tabular}{c|c|c|c}
Level & $J$ & $(J,J)$ & Multiplicity \\ \hline
$ 0 $ & $ (1, 1, -1, -1) $ & $ -2 $ & $ 1 $ \\
$ 1 $ & $ (1, 0, -1, 0) $ & $ -2 $ & $ 1 $ \\
$ 1 $ & $ (0, 1, 0, -1) $ & $ -2 $ & $ 1 $ \\
$ 2 $ & $ (0, 0, 0, 0) $ & $ 0 $ & $ 2 $ \\
$ 3 $ & $ (0, -1, 0, 1) $ & $ -2 $ & $ 1 $ \\
$ 3 $ & $ (-1, 0, 1, 0) $ & $ -2 $ & $ 1 $ \\
$ 4 $ & $ (-1, -1, 1, 1) $ & $ -2 $ & $ 1 $ \\
\end{tabular}
\end{center}
\noindent and we see the structure of an $\mathfrak{su}(3)$ algebra, in full
agreement with expectations from resolution. The linear map
$F:\bZ^4\rightarrow \bZ^2$ from weight junctions to Dynkin labels is
given by $F=
\begin{pmatrix}
0 & 1 & 1 & -1 \\
1 & -1 & -1 & 0
\end{pmatrix}$.  From the level diagram for the roots, we
see that the simple roots in this Weyl chamber are
$\{(1,0,-1,0),(0,1,0,-1)\}$.

We would now like to uncover junctions in the fundamental and
antifundamental representations.  Consider junctions $J\in \bZ^4$ with
$a(J)=\pqvec{-1}{0}$ and $(J,J)=-1$, and also recall that we have
chosen a basis of one-cycles such that $\pi_1 = \pqvec{1}{0}$ and
$\pi_3 = \pqvec{0}{1}$. There are three such junctions. The level
diagram of these junctions is given by
\begin{center}
\begin{tabular}{c|c|c|c}
Level & $J$ & $(J,J)$ & Multiplicity \\ \hline
$0$ & $(0, 1, -1, -1)$ & $-1$ & $1$ \\
$1$ & $(0, 0, -1, 0)$ & $-1$ & $1$ \\
$2$ & $(-1, 0, 0, 0)$ & $-1$ & $1$
\end{tabular}
\end{center}
\noindent and by acting with $F$ on the transpose of the highest weight junction
$(0,1,-1,-1)$ we get (after taking the transpose) $(1,0)$, the Dynkin
labels of the fundamental.  The level diagram of junctions $J \in
\bZ^4$ with $a(J) = \pqvec{1}{0}$ and $(J,J) = -1$ is
\begin{center}
\begin{tabular}{c|c|c|c}
Level & $J$ & $(J,J)$ & Multiplicity \\ \hline
$0$ & $(1, 0, 0, 0)$ & $-1$ & $1$ \\
$1$ & $(0, 0, 1, 0)$ & $-1$ & $1$ \\
$2$ & $(0, -1, 1, 1)$ & $-1$ & $1$
\end{tabular}
\end{center}
\noindent and by applying $F$ to the highest weight junction we see that these
junctions comprise the antifundamental.

In finding the fundamental and the antifundamental, we have identified
highest weight junctions with corresponding Dynkin labels $(1\, 0)$
and $(0\, 1)$; from these we can build \emph{any} representation of
$\mathfrak{su}(3)$. Let us do this in a few examples for the sake of
illustration. One ten dimensional representation has a highest weight
with Dynkin labels $(3,0)$; this is three times the Dynkin labels of
the highest weight of the fundamental, signifying that the $10$ is the
totally symmetric part of the third tensor power of the
fundamental. Taking the highest weight junction $(0,1,-1,-1)$,
multiplying by three to get the highest weight junction of the $10$,
and applying Freudenthal's formula, we obtain the full $10$,
as
\begin{center}
\scalebox{.8}{
\begin{tabular}{c|c|c|c}
Level & $J$ & $(J,J)$ & Multiplicity \\ \hline
$0$ & $(0, 3, -3, -3)$ & $-9$ & $1$ \\
$1$ & $(0, 2, -3, -2)$ & $-5$ & $1$ \\
$2$ & $(0, 1, -3, -1)$ & $-5$ & $1$ \\
$2$ & $(-1, 2, -2, -2)$ & $-5$ & $1$ \\
$3$ & $(0, 0, -3, 0)$ & $-9$ & $1$ \\
$3$ & $(-1, 1, -2, -1)$ & $-3$ & $1$ \\
$4$ & $(-1, 0, -2, 0)$ & $-5$ & $1$ \\
$4$ & $(-2, 1, -1, -1)$ & $-5$ & $1$ \\
$5$ & $(-2, 0, -1, 0)$ & $-5$ & $1$ \\
$6$ & $(-3, 0, 0, 0)$ & $-9$ & $1$
\end{tabular}
}
\end{center}
Alternatively, consider the $27$ dimensional
representation whose highest weight has Dynkin labels $(2,2)$. Adding
two copies of the highest weight junctions of both the fundamental and
antifundamental gives the highest should give the highest weight
junction of the $27$; indeed, applying Freudenthal's formula we see
that it does. These junctions are given by
\begin{center}
\scalebox{.8}{
\begin{tabular}{c|c|c|c}
Level & $J$ & $(J,J)$ & Multiplicity \\ \hline
$0$ & $(2, 2, -2, -2)$ & $-8$ & $1$ \\
$1$ & $(2, 1, -2, -1)$ & $-6$ & $1$ \\
$1$ & $(1, 2, -1, -2)$ & $-6$ & $1$ \\
$2$ & $(2, 0, -2, 0)$ & $-8$ & $1$ \\
$2$ & $(1, 1, -1, -1)$ & $-2$ & $2$ \\
$2$ & $(0, 2, 0, -2)$ & $-8$ & $1$ \\
$3$ & $(1, 0, -1, 0)$ & $-2$ & $2$ \\
$3$ & $(0, 1, 0, -1)$ & $-2$ & $2$ \\
$4$ & $(1, -1, -1, 1)$ & $-6$ & $1$ \\
$4$ & $(0, 0, 0, 0)$ & $0$ & $3$ \\
$4$ & $(-1, 1, 1, -1)$ & $-6$ & $1$ \\
$5$ & $(0, -1, 0, 1)$ & $-2$ & $2$ \\
$5$ & $(-1, 0, 1, 0)$ & $-2$ & $2$ \\
$6$ & $(0, -2, 0, 2)$ & $-8$ & $1$ \\
$6$ & $(-1, -1, 1, 1)$ & $-2$ & $2$ \\
$6$ & $(-2, 0, 2, 0)$ & $-8$ & $1$ \\
$7$ & $(-1, -2, 1, 2)$ & $-6$ & $1$ \\
$7$ & $(-2, -1, 2, 1)$ & $-6$ & $1$ \\
$8$ & $(-2, -2, 2, 2)$ & $-8$ & $1$ \\
\end{tabular}
}
\end{center}
and we could also build any other representation of $\mathfrak{su}(3)$ in a similar manner.

\vspace{.5cm}
\noindent \emph{\textbf{An explicit global to local map.}}
For the sake of illustration, let us briefly show an example global
model into which this local deformation embeds; this is only one of
many, of course.  Consider an elliptic threefold with base
$B=\bP^2$; $f$ and $g$ are sections of $\Gamma(\cO(12))$ and
$\Gamma(\cO(18))$, respectively and thus are degree $12$ and degree
$18$ polynomials in the homogeneous coordinates $(z,t,x_1)$ on
$\bP^2$. Write $f=z^2\, p_{10}$ and $g=z^2\, p_{16}$, with $p_i$ homogenous polynomials of degree $i$. Deform
the manifold by $\wdef{2\ep\, p_{12}}{\ep \,p_{18}}$. Around a general point of $z=0$ the local
geometry is equivalent to that defined by the initial data
(\ref{eq:IVcodim1}) and the deformation (\ref{eq:IVcodim1def}). The gauge algebra associated to the IV singular locus will be then either $\mathfrak{su}(3)$ or generically
    $\mathfrak{sp}(1)\cong \mathfrak{su}(2)$ if there exists  outer monodromy, as in  \cite{Grassi:2013kha}.

\vspace{.5cm}
\noindent \emph{\textbf{$\mathfrak{su}(3)$ gauge algebra from deformation in  a non-Higgsable Cluster.}} As
discussed in section \ref{NHC}, an elliptic fibration over $\bF_3$
necessarily gives rise to a non-Higgsable cluster with a type IV fiber
along the infinity section. It is a simple exercise to write down such
a compact elliptic fibration: in a neighborhood of the locus
$z=0$ of the infinity section $C_\infty$, the elliptic fibration $W_0 \to \mathcal{U} $ is precisely the local elliptic
surface defined by (\ref{eq:IVcodim1}) (see also the vanishing data in the last column of the Table in section \ref{NHC}).   Since this \emph{very
  same} elliptic surface can be obtained via restriction from
geometries which do (see above) or do not admit a global deformation,
the local deformation (\ref{eq:IVcodim1def}) uncovers the same
topological structures in both cases! 

It makes perfect sense physically that the local deformation of the
non-Higgsable cluster uncovers the correct Lie algebraic structure,
even though a global deformation of the geometry does not exist. Physically,
this is just the statement that the geometry still encodes the massless
W-bosons, even though there is no flat direction in the moduli space
which would spontaneously break the gauge symmetry.

\subsection{SU(2) Gauge Symmetry from a Type III Fiber}
\label{sec:su2}
In this section determine the gauge algebra associated to a
deformed type III fiber. From the resolution picture, we know that the algebra
should be $\mathfrak{su}(2)$. However, we will see that this algebra is realized from three
seven-branes, rather than two.

The Weierstrass model we study which exhibits this singular fiber is
\begin{equation}
  f = z \qquad g = z^2.
\end{equation}
and the discriminant is $\Delta = z^3\, (27 z+4)$.
The deformations $\wdef{\ep}{\ep}$ completely break the
degeneracy. The deformed discriminant is
\begin{equation}
\Delta = 4z^3 + 27z^4 + 66\, \ep \, z^2 + 3 \, \ep^2\, (4z+9) + 4\, \ep^3.
\end{equation}
Taking $\ep=.001$ and a neighborhood of $z=0$
the discriminant appears as
\begin{center}
\includegraphics[scale=.8]{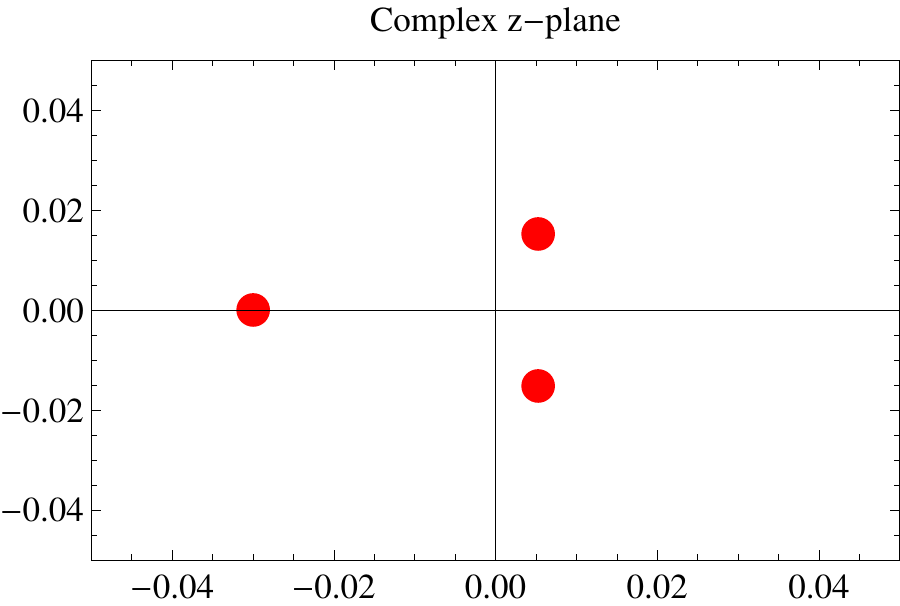}.
\end{center}
and we see that the theory is completely Higgsed, since the
three degenerate defects at $z=0$ are now non-degenerate.
Utilizing a similar technique as in the previous section,
the ordered set of vanishing cycles is determined to be
\begin{equation}
  \label{eq:IIIcodim1}
  Z_{III} = \{ \pi_2,\pi_1,\pi_3\},
\end{equation}
beginning with the leftmost defect and working clockwise around the
origin.  The I-matrix is given by:
\begin{equation}
I = (\cdot,\cdot)=\begin{pmatrix}
-1 & 1/2 & -1/2 \\
1/2 & -1 & 1/2 \\
-1/2 & 1/2 & -1
\end{pmatrix}
\end{equation}
and the roots are $J\in \bZ^3$ such that $(J,J)=-2$ and
$a(J)=\pqvec{0}{0}$, as usual. The root diagram is given by
\begin{center}
\begin{tabular}{c|c|c|c}
Level & $J$ & $(J,J)$ & Multiplicity \\ \hline
$ 0 $ & $ (1, 1, 1) $ & $ -2 $ & $ 1 $ \\
$ 1 $ & $ (0, 0, 0) $ & $ 0 $ & $ 1 $ \\
$ 2 $ & $ (-1, -1, -1) $ & $ -2 $ & $ 1 $ \\
\end{tabular}
\end{center}
which matches an $\mathfrak{su}(2)$ algebra, as expected. Note that unlike the $W_+$ and $W_-$ bosons
of an $\mathfrak{su}(2)$ algebra from an $I_2$ fiber, those arising from a type $III$ fiber are
three pronged string junctions!

For an $\mathfrak{su}(2)$ algebra there is no way to distinguish a fundamental
from an antifundamental. However, there are two simple sets of
junctions which can be distinguished, but nevertheless map to $(1)$
under the map $F$, which here is given by
$F=\begin{pmatrix}1&0&1\end{pmatrix}$. These are the sets
\begin{align}
  \left\{ J \,\, |\,\, (J,J) = -1 \,\, \text{and} \,\, a(J) = \pqvec{1}{1} \right\} &= \left\{(0, 1, 1),(-1,0, 0) \right\}  \nonumber \\
  \left\{ J \,\, |\,\, (J,J) = -1 \,\, \text{and} \,\, a(J) =
    \pqvec{-1}{0}\right\} &= \left\{(1, 0, 0),(0, -1, -1) \right\}
\end{align}
We see that there is geometric data which differentiates between two
different realizations of doublets, despite being the same Lie algebra
representation; this phenomenon and associated physics implications
were discussed\footnote{There it was related to
additional constraints on $\mathfrak{su}(2)$ gauge theories (in another context)
which are necessary and sufficient for anomaly cancellation in
nucleated D-brane theories.} in \cite{Halverson:2013ska}.

Of course, from this data higher dimensional representations can be built up as
in section \ref{sec:su3}. For example, taking the junction $(0,1,1)$ and multiplying
by four should give junctions in the fourth symmetric tensor power of the $2$ of $\mathfrak{su}(2)$;
that is, the $5$. Indeed, Freudenthal's formula applied to $(0,4,4)$ gives
\begin{center}
\begin{tabular}{c|c|c|c}
Level & $J$ & $(J,J)$ & Multiplicity \\ \hline
$0$ & $(0, 4, 4)$ & $-16$ & $1$ \\
$1$ & $(-1, 3, 3)$ & $-10$ & $1$ \\
$2$ & $(-2, 2, 2)$ & $-8$ & $1$ \\
$3$ & $(-3, 1, 1)$ & $-10$ & $1$ \\
$4$ & $(-4, 0, 0)$ & $-16$ & $1$ \\
\end{tabular}
\end{center}
which are the junctions in the $5$ of $\mathfrak{su}(2)$.

\subsection{No Gauge Symmetry from a Type II Fiber}
\label{sec:nogaugefromtypeII}
Finally, we would like to briefly discuss type II fibers. These are interesting
because though they are in Kodaira's classification of singular fibers
in codimension one, they are known to carry no gauge algebra, despite having
$\Delta \sim z^2$. We will see that deformation quite easily recovers this fact.

The local Weierstrass model we study which realized this singular
fiber is defined by
\begin{equation}
  f=z \qquad g=z
\end{equation}
with $\Delta = z^2 \, (4 z+27)$.
Consider the completely Higgsing $fg$-deformation
$\wdef{2\,\ep}{\ep}$. The deformed discriminant is
\begin{equation}
  \Delta = 4 z^3+ 27 z^2 +  \epsilon \, (24  z^2  + 54 z) + \epsilon^2\,( 48 z + 27)  + 32\, \epsilon ^3
\end{equation}
Taking $\ep=2$, the the discriminant in a neighborhood of
$z=0$ is given by
\begin{center}
\includegraphics[scale=.8]{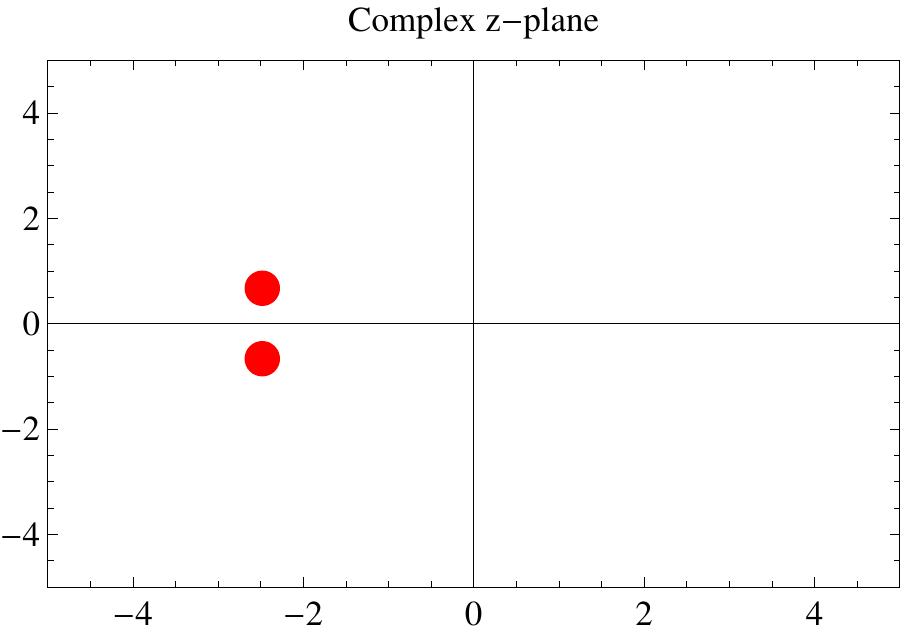}.
\end{center}
and taking straight line paths of approach from the origin, the vanishing
cycles are determined to be
\begin{equation}
Z_{II} = \{\pi_3,\pi_1\},
\end{equation}
where $\pi_3$ ($\pi_1$) is associated to the discriminant component in
the upper (lower) left quadrant. The $I$-matrix is given by
\begin{equation}
I = (\cdot,\cdot)=\begin{pmatrix}
-1 & -1/2 \\
-1/2 & -1
\end{pmatrix}
\end{equation}
With these vanishing cycles and this $I$-matrix, it is easy to see
that there are no junctions $J$ with $(J,J)=-2$ and $a(J) =
\pqvec{0}{0}$; i.e. there are no two-manifolds which would be roots,
and therefore there is no gauge algebra. From the non-trivial
structure of vanishing cycles, however, codimension two collisons of a
type II fiber with another fiber could yield an interesting structure
of matter representations.

\clearpage
\section{Conclusions}
\label{sec:conclusions}

In this paper we have studied non-abelian gauge symmetry in F-theory
compactifications on an elliptically fibered Calabi-Yau variety
$X$. We have emphasized that the technique we utilize --- singularity
deformation by movement in the complex structure moduli space of $X$
--- is an ideal technique for studying the structure of non-abelian
gauge theories in F-theory, because this branch of the moduli space
exists in both F-theory and its defining M-theory compactification,
and furthermore encompasses and extends the descriptions of gauge
theories in compactifications with a heterotic dual or a weakly
coupled type IIb limit.

In the second part of the introduction we reviewed the definition of
F-theory in terms of M-theory, the drawbacks of the M-theory Coulomb
branch of singularity resolution as a technique to study F-theory, and
the advantages of utilizing complex structure deformation.  These,
along with the discussions of section \ref{sec:deformation physics}
lead to the following important points, some
of which are known:
\begin{itemize}
\item The resolution of singular fibers does not describe the
  spontaneous breaking\\ of any gauge symmetry in F-theory.
\item There can be an M-theory Coulomb branch which lifts to F-theory. \\
  However, it is the one of deformation, not resolution; the latter does
  not lift.
\item Complex structure moduli determine the expectation values of
  scalar fields in F-theory. \\ In $d=4$ these are the scalar fields
  in chiral multiplets.
\item Singularity resolution cannot describe massive gauge states in
  F-theory.
\item The gauge states in F-theory arising from singular fibers --
  which are the ones typically studied in the literature -- \emph{are}
  string junctions, and in the zero size limit if they are massless.
\end{itemize}
These physical points motivate the approach of the rest of the work.
In section \ref{sec:glimpse} we determined the Lie algebraic structure
of a spontaneously broken global model in a simple example.  In
section \ref{sec:deformations} we discussed the physics of
deformations and presented a systematic discussion of complex
structure deformations which are useful for studying gauge theoretic
data. We also show how to use a local deformation, together with a
global geometry analysis, to deduce global information about gauge
groups and matter representations, even in cases where no global
deformation is possible; i.e. the case of non-Higgsable clusters.  In
section \ref{sec:less studied fibers} we presented three new examples
of using deformation theory to read off gauge theoretic data; these
are for the less studied type $IV$, $III$, and $II$ fibers, which
realize $\mathfrak{su}(3)$, $\mathfrak{su}(2)$ and $\emptyset$ gauge theories,
respectively. Interestingly, the $\mathfrak{su}(3)$ and $\mathfrak{su}(2)$ theories are
realized by states ending on four and three seven-branes,
respectively, in contrast with the $D7$-brane case. Some number of
roots of these algebras arise from junctions with more than two
prongs, providing additional direct evidence that they do not have a
weakly coupled type IIb description.

It is natural to expect that certain aspects of the physics of
four-dimensional F-theory compactifications will be elucidated by
utilizing the branch of the moduli space that exists in the theory.
Some are immediately clear; e.g.  $G_4$-fluxes in $d=3$
$\cN=2$ M-theory compactifications which lift to Lorentz-invariant
configurations in F-theory have precisely one leg along the fiber
\cite{Dasgupta:1999ss}. What physical objects are these $G_4$-fluxes
``along''?  They should be along the $M2$-branes of the deformation
picture, which wrap precisely one dimension of the elliptic
fiber. This simple physical point is obscured in the resolution
picture, where $M2$-branes states wrap two-cycles which are entirely in
the resolved fiber, and it is not clear what the necessary ``one leg along
the fiber'' flux condition means physically.

We believe that many other interesting aspects of F-theory
compactifications will be better understood in the deformation
picture. We leave such investigations to future work.

\vspace{1cm}
\noindent \textbf{Acknowledgments.}  It is a pleasure to thank Lara
Anderson, Mirjam Cveti\v{c}, I\~ naki Garc\' ia-Etxebarria, Tom
Hartman, Denis Klevers, Peter Koroteev, Liam McAllister, Paul McGuirk,
Joe Polchinski, Wati Taylor and especially David R. Morrison for
useful conversations. We would also like to thank the
  referee for
  many valuable suggestions. J.H. thanks J.L. Halverson for her kind
support and constant encouragement. He is supported by the National
Science Foundation under Grant No. PHY11-25915. J.L.S. is supported by
DARPA, fund no. 553700 and is the Class of 1939 Professor in the School
of Arts and Sciences of the University of Pennsylvania and gratefully
acknowledges the generosity of the Class of 1939. A.G. is supported by
NSF Research Training Group Grant DMS-0636606.

\appendix

\section{Freely Available Computer Packages}
\label{sec:code}

In completing this work, and also our previous work \cite{Grassi:2013kha},
we have written computer codes to perform many computations. These
codes are publcly available at
\begin{center}
\url{http://www.jhhalverson.com/deformations}
\end {center}
and we hope that the interested reader finds them useful.  There you
can find codes for all of the examples in this paper. We have tried to
make them clear via including these examples and also comments,
but encourage questions on the codes if these are not sufficient.

We have utilized two types of codes. The first are Mathematica
notebooks which make it simple to read off vanishing cycles in
deformed geometries.  This data can then be fed into our package {\tt
  py-junctions} which can perform a number of Lie algebraic
computations.  Though it is written in Python, it is best executed
through a SAGE terminal, since it utilizes packages which are
automatically included in SAGE.

\bibliographystyle{utphys}
\bibliography{refs}

\providecommand{\href}[2]{#2}\begingroup\raggedright\begin{thebibliography}{10}

\bibitem{Vafa:1996xn}
C.~Vafa, ``{Evidence for F-Theory},'' {\em Nucl. Phys.} {\bf B469} (1996)
  403--418,
\href{http://www.arXiv.org/abs/hep-th/9602022}{{\tt hep-th/9602022}}.
%%CITATION = HEP-TH/9602022;%%.

\bibitem{Beasley:2008dc}
C.~Beasley, J.~J. Heckman, and C.~Vafa, ``{Guts and Exceptional Branes in
  F-Theory - I},'' {\em JHEP} {\bf 01} (2009) 058,
\href{http://www.arXiv.org/abs/0802.3391}{{\tt 0802.3391}}.
%%CITATION = 0802.3391;%%.

\bibitem{Donagi:2008ca}
R.~Donagi and M.~Wijnholt, ``{Model Building with F-Theory},''
\href{http://www.arXiv.org/abs/0802.2969}{{\tt 0802.2969}}.
%%CITATION = 0802.2969;%%.

\bibitem{Andreas:2009uf}
B.~Andreas and G.~Curio, ``{From Local to Global in F-Theory Model Building},''
  {\em J.Geom.Phys.} {\bf 60} (2010) 1089--1102,
\href{http://www.arXiv.org/abs/0902.4143}{{\tt 0902.4143}}.
%%CITATION = ARXIV:0902.4143;%%.

\bibitem{Marsano:2009ym}
J.~Marsano, N.~Saulina, and S.~Sch{\"a}fe{r-Na}meki, ``{F-Theory
  Compactifications for Supersymmetric Guts},'' {\em JHEP} {\bf 08} (2009) 030,
\href{http://www.arXiv.org/abs/0904.3932}{{\tt 0904.3932}}.
%%CITATION = 0904.3932;%%.

\bibitem{Collinucci:2009uh}
A.~Collinucci, ``{New F-Theory Lifts Ii: Permutation Orientifolds and Enhanced
  Singularities},'' {\em JHEP} {\bf 04} (2010) 076,
\href{http://www.arXiv.org/abs/0906.0003}{{\tt 0906.0003}}.
%%CITATION = 0906.0003;%%.

\bibitem{Blumenhagen:2009up}
R.~Blumenhagen, T.~W. Grimm, B.~Jurke, and T.~Weigand, ``{F-Theory Uplifts and
  Guts},'' {\em JHEP} {\bf 09} (2009) 053,
\href{http://www.arXiv.org/abs/0906.0013}{{\tt 0906.0013}}.
%%CITATION = 0906.0013;%%.

\bibitem{Marsano:2009gv}
J.~Marsano, N.~Saulina, and S.~Sch{\"a}fe{r-Na}meki, ``{Monodromies, Fluxes,
  and Compact Three-Generation F-Theory Guts},'' {\em JHEP} {\bf 08} (2009)
  046,
\href{http://www.arXiv.org/abs/0906.4672}{{\tt 0906.4672}}.
%%CITATION = 0906.4672;%%.

\bibitem{Blumenhagen:2009yv}
R.~Blumenhagen, T.~W. Grimm, B.~Jurke, and T.~Weigand, ``{Global F-Theory
  Guts},'' {\em Nucl. Phys.} {\bf B829} (2010) 325--369,
\href{http://www.arXiv.org/abs/0908.1784}{{\tt 0908.1784}}.
%%CITATION = 0908.1784;%%.

\bibitem{Marsano:2009wr}
J.~Marsano, N.~Saulina, and S.~Sch{\"a}fe{r-Na}meki, ``{Compact F-Theory Guts
  with $U(1)_{PQ}$},'' {\em JHEP} {\bf 04} (2010) 095,
\href{http://www.arXiv.org/abs/0912.0272}{{\tt 0912.0272}}.
%%CITATION = 0912.0272;%%.

\bibitem{Grimm:2009yu}
T.~W. Grimm, S.~Krause, and T.~Weigand, ``{F-Theory GUT Vacua on Compact
  Calabi-Yau Fourfolds},'' {\em JHEP} {\bf 07} (2010) 037,
\href{http://www.arXiv.org/abs/0912.3524}{{\tt 0912.3524}}.
%%CITATION = 0912.3524;%%.

\bibitem{Cvetic:2010rq}
M.~Cveti{\v c}, I.~Garc{\'\i a-}Etxebarria, and J.~Halverson, ``{Global
  F-Theory Models: Instantons and Gauge Dynamics},'' {\em JHEP} {\bf 1101}
  (2011) 073,
\href{http://www.arXiv.org/abs/1003.5337}{{\tt 1003.5337}}.
%%CITATION = ARXIV:1003.5337;%%.

\bibitem{Chen:2010ts}
C.-M. Chen, J.~Knapp, M.~Kreuzer, and C.~Mayrhofer, ``{Global SO(10) F-Theory
  Guts},'' {\em JHEP} {\bf 10} (2010) 057,
\href{http://www.arXiv.org/abs/1005.5735}{{\tt 1005.5735}}.
%%CITATION = 1005.5735;%%.

\bibitem{Chen:2010tp}
C.-M. Chen and Y.-C. Chung, ``{Flipped $SU(5)$ Guts from $E_{8}$ Singularities
  in F-Theory},'' {\em JHEP} {\bf 03} (2011) 049,
\href{http://www.arXiv.org/abs/1005.5728}{{\tt 1005.5728}}.
%%CITATION = 1005.5728;%%.

\bibitem{Chung:2010bn}
Y.-C. Chung, ``{On Global Flipped $SU(5)$ Guts in F-Theory},'' {\em JHEP} {\bf
  03} (2011) 126,
\href{http://www.arXiv.org/abs/1008.2506}{{\tt 1008.2506}}.
%%CITATION = 1008.2506;%%.

\bibitem{Chen:2010tg}
C.-M. Chen and Y.-C. Chung, ``{On F-Theory $E_{6}$ Guts},'' {\em JHEP} {\bf 03}
  (2011) 129,
\href{http://www.arXiv.org/abs/1010.5536}{{\tt 1010.5536}}.
%%CITATION = 1010.5536;%%.

\bibitem{Knapp:2011wk}
J.~Knapp, M.~Kreuzer, C.~Mayrhofer, and N.-O. Walliser, ``{Toric Construction
  of Global F-Theory Guts},'' {\em JHEP} {\bf 03} (2011) 138,
\href{http://www.arXiv.org/abs/1101.4908}{{\tt 1101.4908}}.
%%CITATION = 1101.4908;%%.

\bibitem{Knapp:2011ip}
J.~Knapp and M.~Kreuzer, ``{Toric Methods in F-Theory Model Building},'' {\em
  Adv.High Energy Phys.} {\bf 2011} (2011) 513436,
\href{http://www.arXiv.org/abs/1103.3358}{{\tt 1103.3358}}.
%%CITATION = ARXIV:1103.3358;%%.

\bibitem{Marsano:2012yc}
J.~Marsano, H.~Clemens, T.~Pantev, S.~Raby, and H.-H. Tseng, ``{A Global
  $SU(5)$ F-Theory Model with Wilson Line Breaking},''
\href{http://www.arXiv.org/abs/1206.6132}{{\tt 1206.6132}}.
%%CITATION = ARXIV:1206.6132;%%.

\bibitem{Grimm:2010ez}
T.~W. Grimm and T.~Weigand, ``{On Abelian Gauge Symmetries and Proton Decay in
  Global F- Theory Guts},'' {\em Phys. Rev.} {\bf D82} (2010) 086009,
\href{http://www.arXiv.org/abs/1006.0226}{{\tt 1006.0226}}.
%%CITATION = 1006.0226;%%.

\bibitem{Dolan:2011iu}
M.~J. Dolan, J.~Marsano, N.~Saulina, and S.~Sch{\"a}fe{r-Na}meki, ``{F-Theory
  Guts with U(1) Symmetries: Generalities and Survey},'' {\em Phys.Rev.} {\bf
  D84} (2011) 066008,
\href{http://www.arXiv.org/abs/1102.0290}{{\tt 1102.0290}}.
%%CITATION = ARXIV:1102.0290;%%.

\bibitem{Marsano:2011nn}
J.~Marsano, N.~Saulina, and S.~Sch{\"a}fe{r-Na}meki, ``{On G-Flux, M5
  Instantons, and U(1)S in F-Theory},''
\href{http://www.arXiv.org/abs/1107.1718}{{\tt 1107.1718}}.
%%CITATION = 1107.1718;%%.

\bibitem{Grimm:2011tb}
T.~W. Grimm, M.~Kerstan, E.~Palti, and T.~Weigand, ``{Massive Abelian Gauge
  Symmetries and Fluxes in F-Theory},'' {\em JHEP} {\bf 1112} (2011) 004,
\href{http://www.arXiv.org/abs/1107.3842}{{\tt 1107.3842}}.
%%CITATION = ARXIV:1107.3842;%%.

\bibitem{Morrison:2012ei}
D.~R. Morrison and D.~S. Park, ``{F-Theory and the Mordell-Weil Group of
  Elliptically-Fibered Calabi-Yau Threefolds},''
\href{http://www.arXiv.org/abs/1208.2695}{{\tt 1208.2695}}.
%%CITATION = ARXIV:1208.2695;%%.

\bibitem{Mayrhofer:2012zy}
C.~Mayrhofer, E.~Palti, and T.~Weigand, ``{U(1) Symmetries in F-Theory Guts
  with Multiple Sections},'' {\em JHEP} {\bf 1303} (2013) 098,
\href{http://www.arXiv.org/abs/1211.6742}{{\tt 1211.6742}}.
%%CITATION = ARXIV:1211.6742;%%.

\bibitem{Cvetic:2013nia}
M.~Cveti{\v c}, D.~Klevers, and H.~Piragua, ``{F-Theory Compactifications with
  Multiple U(1)-Factors: Constructing Elliptic Fibrations with Rational
  Sections},''
\href{http://www.arXiv.org/abs/1303.6970}{{\tt 1303.6970}}.
%%CITATION = ARXIV:1303.6970;%%.

\bibitem{Cvetic:2013jta}
M.~Cveti{\v c}, D.~Klevers, and H.~Piragua, ``{F-Theory Compactifications with
  Multiple U(1)-Factors: Addendum},'' {\em JHEP} {\bf 1312} (2013) 056,
\href{http://www.arXiv.org/abs/1307.6425}{{\tt 1307.6425}}.
%%CITATION = ARXIV:1307.6425;%%.

\bibitem{Cvetic:2013qsa}
M.~Cveti{\v c}, D.~Klevers, H.~Piragua, and P.~Song, ``{Elliptic Fibrations
  with Rank Three Mordell-Weil Group: F-Theory with U(1) $\times$ U(1) $\times$
  U(1) Gauge Symmetry},''
\href{http://www.arXiv.org/abs/1310.0463}{{\tt 1310.0463}}.
%%CITATION = ARXIV:1310.0463;%%.

\bibitem{Borchmann:2013jwa}
J.~Borchmann, C.~Mayrhofer, E.~Palti, and T.~Weigand, ``{Elliptic Fibrations
  for $SU(5)$ $\times$ U(1) $\times$ U(1) F-Theory Vacua},''
\href{http://www.arXiv.org/abs/1303.5054}{{\tt 1303.5054}}.
%%CITATION = ARXIV:1303.5054;%%.

\bibitem{Borchmann:2013hta}
J.~Borchmann, C.~Mayrhofer, E.~Palti, and T.~Weigand, ``{$SU(5)$ Tops with
  Multiple U(1)S in F-Theory},''
\href{http://www.arXiv.org/abs/1307.2902}{{\tt 1307.2902}}.
%%CITATION = ARXIV:1307.2902;%%.

\bibitem{Braun:2013nqa}
V.~Braun, T.~W. Grimm, and J.~Keitel, ``{Geometric Engineering in Toric
  F-Theory and Guts with U(1) Gauge Factors},'' {\em JHEP} {\bf 1312} (2013)
  069,
\href{http://www.arXiv.org/abs/1306.0577}{{\tt 1306.0577}}.
%%CITATION = ARXIV:1306.0577;%%.

\bibitem{Braun:2013yti}
V.~Braun, T.~W. Grimm, and J.~Keitel, ``{New Global F-Theory Guts with U(1)
  Symmetries},'' {\em JHEP} {\bf 1309} (2013) 154,
\href{http://www.arXiv.org/abs/1302.1854}{{\tt 1302.1854}}.
%%CITATION = ARXIV:1302.1854;%%.

\bibitem{Braun:2014nva}
A.~P. Braun, A.~Collinucci, and R.~Valandro, ``{The Fate of U(1)'s at Strong
  Coupling in F-Theory},''
\href{http://www.arXiv.org/abs/1402.4054}{{\tt 1402.4054}}.
%%CITATION = ARXIV:1402.4054;%%.

\bibitem{Blumenhagen:2010ja}
R.~Blumenhagen, A.~Collinucci, and B.~Jurke, ``{On Instanton Effects in
  F-Theory},'' {\em JHEP} {\bf 08} (2010) 079,
\href{http://www.arXiv.org/abs/1002.1894}{{\tt 1002.1894}}.
%%CITATION = 1002.1894;%%.

\bibitem{Donagi:2010pd}
R.~Donagi and M.~Wijnholt, ``{Msw Instantons},''
\href{http://www.arXiv.org/abs/1005.5391}{{\tt 1005.5391}}.
%%CITATION = ARXIV:1005.5391;%%.

\bibitem{Grimm:2011dj}
T.~W. Grimm, M.~Kerstan, E.~Palti, and T.~Weigand, ``{On Fluxed Instantons and
  Moduli Stabilisation in IIB Orientifolds and F-Theory},''
\href{http://www.arXiv.org/abs/1105.3193}{{\tt 1105.3193}}.
%%CITATION = 1105.3193;%%.

\bibitem{Cvetic:2011gp}
M.~Cveti{\v c}, I.~Garcia~Etxebarria, and J.~Halverson, ``{Three Looks at
  Instantons in F-theory -- New Insights from Anomaly Inflow, String Junctions
  and Heterotic Duality},'' {\em JHEP} {\bf 1111} (2011) 101,
\href{http://www.arXiv.org/abs/1107.2388}{{\tt 1107.2388}}.
%%CITATION = ARXIV:1107.2388;%%.

\bibitem{Bianchi:2011qh}
M.~Bianchi, A.~Collinucci, and L.~Martucci, ``{Magnetized E3-Brane Instantons
  in F-Theory},'' {\em JHEP} {\bf 1112} (2011) 045,
\href{http://www.arXiv.org/abs/1107.3732}{{\tt 1107.3732}}.
%%CITATION = ARXIV:1107.3732;%%.

\bibitem{Kerstan:2012cy}
M.~Kerstan and T.~Weigand, ``{Fluxed M5-Instantons in F-Theory},''
\href{http://www.arXiv.org/abs/1205.4720}{{\tt 1205.4720}}.
%%CITATION = ARXIV:1205.4720;%%.

\bibitem{Cvetic:2012ts}
M.~Cveti{\v c}, R.~Donagi, J.~Halverson, and J.~Marsano, ``{On Seven-Brane
  Dependent Instanton Prefactors in F-Theory},'' {\em JHEP} {\bf 1211} (2012)
  004,
\href{http://www.arXiv.org/abs/1209.4906}{{\tt 1209.4906}}.
%%CITATION = ARXIV:1209.4906;%%.

\bibitem{Bianchi:2012kt}
M.~Bianchi, G.~Inverso, and L.~Martucci, ``{Brane Instantons and Fluxes in
  F-Theory},''
\href{http://www.arXiv.org/abs/1212.0024}{{\tt 1212.0024}}.
%%CITATION = ARXIV:1212.0024;%%.

\bibitem{Grassi:2000we}
A.~Grassi and D.~R. Morrison, ``{Group Representations and the Euler
  Characteristic of Elliptically Fibered Calabi-Yau Threefolds},''
\href{http://www.arXiv.org/abs/math/0005196}{{\tt math/0005196}}.
%%CITATION = MATH/0005196;%%.

\bibitem{Grassi:2011hq}
A.~Grassi and D.~R. Morrison, ``{Anomalies and the Euler Characteristic of
  Elliptic Calabi-Yau Threefolds},''
\href{http://www.arXiv.org/abs/1109.0042}{{\tt 1109.0042}}.
%%CITATION = ARXIV:1109.0042;%%.

\bibitem{Morrison:2011mb}
D.~R. Morrison and W.~Taylor, ``{Matter and Singularities},'' {\em JHEP} {\bf
  1201} (2012) 022,
\href{http://www.arXiv.org/abs/1106.3563}{{\tt 1106.3563}}.
%%CITATION = ARXIV:1106.3563;%%.

\bibitem{Grassi:2013kha}
A.~Grassi, J.~Halverson, and J.~L. Shaneson, ``{Matter from Geometry without
  Resolution},''
\href{http://www.arXiv.org/abs/1306.1832}{{\tt 1306.1832}}.
%%CITATION = ARXIV:1306.1832;%%.

\bibitem{Lawrie:2012gg}
C.~Lawrie and S.~Schäfer-Nameki, ``{The Tate Form on Steroids: Resolution and
  Higher Codimension Fibers},'' {\em JHEP} {\bf 1304} (2013) 061,
\href{http://www.arXiv.org/abs/1212.2949}{{\tt 1212.2949}}.
%%CITATION = ARXIV:1212.2949;%%.

\bibitem{Hayashi:2013lra}
H.~Hayashi, C.~Lawrie, and S.~Schafer-Nameki, ``{Phases, Flops and F-theory:
  SU(5) Gauge Theories},'' {\em JHEP} {\bf 1310} (2013) 046,
\href{http://www.arXiv.org/abs/1304.1678}{{\tt 1304.1678}}.
%%CITATION = ARXIV:1304.1678;%%.

\bibitem{Hayashi:2014kca}
H.~Hayashi, C.~Lawrie, D.~R. Morrison, and S.~Sch{\"a}fer-Nameki, ``{Box Graphs
  and Singular Fibers},''
\href{http://www.arXiv.org/abs/1402.2653}{{\tt 1402.2653}}.
%%CITATION = ARXIV:1402.2653;%%.

\bibitem{Bonora:2010bu}
L.~Bonora and R.~Savelli, ``{Non-Simply-Laced Lie Algebras via F Theory
  Strings},'' {\em JHEP} {\bf 1011} (2010) 025,
\href{http://www.arXiv.org/abs/1007.4668}{{\tt 1007.4668}}.
%%CITATION = ARXIV:1007.4668;%%.

\bibitem{Mizoguchi:2014gva}
S.~Mizoguchi, ``{F-theory Family Unification},''
\href{http://www.arXiv.org/abs/1403.7066}{{\tt 1403.7066}}.
%%CITATION = ARXIV:1403.7066;%%.

\bibitem{Marsano:2010ix}
J.~Marsano, N.~Saulina, and S.~Sch{\"a}fe{r-Na}meki, ``{A Note on G-Fluxes for
  F-Theory Model Building},'' {\em JHEP} {\bf 11} (2010) 088,
\href{http://www.arXiv.org/abs/1006.0483}{{\tt 1006.0483}}.
%%CITATION = 1006.0483;%%.

\bibitem{Collinucci:2010gz}
A.~Collinucci and R.~Savelli, ``{On Flux Quantization in F-Theory},'' {\em
  JHEP} {\bf 1202} (2012) 015,
\href{http://www.arXiv.org/abs/1011.6388}{{\tt 1011.6388}}.
%%CITATION = ARXIV:1011.6388;%%.

\bibitem{Braun:2011zm}
A.~P. Braun, A.~Collinucci, and R.~Valandro, ``{G-Flux in F-Theory and
  Algebraic Cycles},'' {\em Nucl.Phys.} {\bf B856} (2012) 129--179,
\href{http://www.arXiv.org/abs/1107.5337}{{\tt 1107.5337}}.
%%CITATION = ARXIV:1107.5337;%%.

\bibitem{Marsano:2011hv}
J.~Marsano and S.~Sch{\"a}fe{r-Na}meki, ``{Yukawas, G-Flux, and Spectral Covers
  from Resolved Calabi- Yau's},'' {\em JHEP} {\bf 11} (2011) 098,
\href{http://www.arXiv.org/abs/1108.1794}{{\tt 1108.1794}}.
%%CITATION = 1108.1794;%%.

\bibitem{Krause:2011xj}
S.~Krause, C.~Mayrhofer, and T.~Weigand, ``{$G_4$ Flux, Chiral Matter and
  Singularity Resolution in F-Theory Compactifications},'' {\em Nucl.Phys.}
  {\bf B858} (2012) 1--47,
\href{http://www.arXiv.org/abs/1109.3454}{{\tt 1109.3454}}.
%%CITATION = ARXIV:1109.3454;%%.

\bibitem{Grimm:2011fx}
T.~W. Grimm and H.~Hayashi, ``{F-Theory Fluxes, Chirality and Chern-Simons
  Theories},''
\href{http://www.arXiv.org/abs/1111.1232}{{\tt 1111.1232}}.
%%CITATION = 1111.1232;%%.

\bibitem{Braun:2012nk}
A.~P. Braun, A.~Collinucci, and R.~Valandro, ``{Algebraic Description of G-Flux
  in F-Theory: New Techniques for F-Theory Phenomenology},''
\href{http://www.arXiv.org/abs/1202.5029}{{\tt 1202.5029}}.
%%CITATION = ARXIV:1202.5029;%%.

\bibitem{Kuntzler:2012bu}
M.~Kuntzler and S.~Sch{\"a}fe{r-Na}meki, ``{G-Flux and Spectral Divisors},''
\href{http://www.arXiv.org/abs/1205.5688}{{\tt 1205.5688}}.
%%CITATION = ARXIV:1205.5688;%%.

\bibitem{Krause:2012he}
S.~Krause, C.~Mayrhofer, and T.~Weigand, ``{Gauge Fluxes in F-Theory and Type
  IIB Orientifolds},'' {\em JHEP} {\bf 1208} (2012) 119,
\href{http://www.arXiv.org/abs/1202.3138}{{\tt 1202.3138}}.
%%CITATION = ARXIV:1202.3138;%%.

\bibitem{Collinucci:2012as}
A.~Collinucci and R.~Savelli, ``{On Flux Quantization in F-Theory Ii: Unitary
  and Symplectic Gauge Groups},'' {\em JHEP} {\bf 1208} (2012) 094,
\href{http://www.arXiv.org/abs/1203.4542}{{\tt 1203.4542}}.
%%CITATION = ARXIV:1203.4542;%%.

\bibitem{Marsano:2012bf}
J.~Marsano, N.~Saulina, and S.~Sch{\"a}fe{r-Na}meki, ``{Global Gluing and
  G-Flux},''
\href{http://www.arXiv.org/abs/1211.1097}{{\tt 1211.1097}}.
%%CITATION = ARXIV:1211.1097;%%.

\bibitem{Intriligator:2012ue}
K.~Intriligator, H.~Jockers, P.~Mayr, D.~R. Morrison, and M.~R. Plesser,
  ``{Conifold Transitions in M-theory on Calabi-Yau Fourfolds with Background
  Fluxes},''
\href{http://www.arXiv.org/abs/1203.6662}{{\tt 1203.6662}}.
%%CITATION = ARXIV:1203.6662;%%.

\bibitem{Anderson:2013rka}
L.~B. Anderson, J.~J. Heckman, and S.~Katz, ``{T-Branes and Geometry},''
\href{http://www.arXiv.org/abs/1310.1931}{{\tt 1310.1931}}.
%%CITATION = ARXIV:1310.1931;%%.

\bibitem{Kumar:2009ac}
V.~Kumar, D.~R. Morrison, and W.~Taylor, ``{Mapping 6D ${\mathcal{N}}\!=1$
  Supergravities to F-Theory},'' {\em JHEP} {\bf 02} (2010) 099,
\href{http://www.arXiv.org/abs/0911.3393}{{\tt 0911.3393}}.
%%CITATION = 0911.3393;%%.

\bibitem{Kumar:2010ru}
V.~Kumar, D.~R. Morrison, and W.~Taylor, ``{Global Aspects of the Space of 6D
  ${\mathcal{N}}\!=1$ Supergravities},'' {\em JHEP} {\bf 1011} (2010) 118,
\href{http://www.arXiv.org/abs/1008.1062}{{\tt 1008.1062}}.
%%CITATION = ARXIV:1008.1062;%%.

\bibitem{Kumar:2010am}
V.~Kumar, D.~S. Park, and W.~Taylor, ``{6D Supergravity without Tensor
  Multiplets},'' {\em JHEP} {\bf 1104} (2011) 080,
\href{http://www.arXiv.org/abs/1011.0726}{{\tt 1011.0726}}.
%%CITATION = ARXIV:1011.0726;%%.

\bibitem{Morrison:2012np}
D.~R. Morrison and W.~Taylor, ``{Classifying Bases for 6D F-Theory Models},''
  {\em Central Eur.J.Phys.} {\bf 10} (2012) 1072--1088,
\href{http://www.arXiv.org/abs/1201.1943}{{\tt 1201.1943}}.
%%CITATION = ARXIV:1201.1943;%%.

\bibitem{Braun:2014oya}
V.~Braun and D.~R. Morrison, ``{F-Theory on Genus-One Fibrations},''
\href{http://www.arXiv.org/abs/1401.7844}{{\tt 1401.7844}}.
%%CITATION = ARXIV:1401.7844;%%.

\bibitem{Heckman:2013pva}
J.~J. Heckman, D.~R. Morrison, and C.~Vafa, ``{On the Classification of 6D
  SCFTs and Generalized ADE Orbifolds},''
\href{http://www.arXiv.org/abs/1312.5746}{{\tt 1312.5746}}.
%%CITATION = ARXIV:1312.5746;%%.

\bibitem{Affleck:1982as}
I.~Affleck, J.~A. Harvey, and E.~Witten, ``{Instantons and (Super)Symmetry
  Breaking in (2+1)-Dimensions},'' {\em Nucl.Phys.} {\bf B206} (1982)
413.
%%CITATION = NUPHA,B206,413;%%.

\bibitem{Aharony:1997bx}
O.~Aharony, A.~Hanany, K.~A. Intriligator, N.~Seiberg, and M.~Strassler,
  ``{Aspects of ${\mathcal{N}}\!=2$ Supersymmetric Gauge Theories in
  Three-Dimensions},'' {\em Nucl.Phys.} {\bf B499} (1997) 67--99,
\href{http://www.arXiv.org/abs/hep-th/9703110}{{\tt hep-th/9703110}}.
%%CITATION = HEP-TH/9703110;%%.

\bibitem{deBoer:1997kr}
J.~de~Boer, K.~Hori, and Y.~Oz, ``{Dynamics of ${\mathcal{N}}\!=2$
  Supersymmetric Gauge Theories in Three-Dimensions},'' {\em Nucl.Phys.} {\bf
  B500} (1997) 163--191,
\href{http://www.arXiv.org/abs/hep-th/9703100}{{\tt hep-th/9703100}}.
%%CITATION = HEP-TH/9703100;%%.

\bibitem{Katz:1996th}
S.~H. Katz and C.~Vafa, ``{Geometric Engineering of ${\mathcal{N}}\!=1$ Quantum
  Field Theories},'' {\em Nucl.Phys.} {\bf B497} (1997) 196--204,
\href{http://www.arXiv.org/abs/hep-th/9611090}{{\tt hep-th/9611090}}.
%%CITATION = HEP-TH/9611090;%%.

\bibitem{Morrison:1996na}
D.~R. Morrison and C.~Vafa, ``{Compactifications of F-Theory on Calabi--Yau
  Threefolds -- I},'' {\em Nucl. Phys.} {\bf B473} (1996) 74--92,
\href{http://www.arXiv.org/abs/hep-th/9602114}{{\tt hep-th/9602114}}.
%%CITATION = HEP-TH/9602114;%%.

\bibitem{Morrison:1996pp}
D.~R. Morrison and C.~Vafa, ``{Compactifications of F-Theory on Calabi--Yau
  Threefolds -- II},'' {\em Nucl. Phys.} {\bf B476} (1996) 437--469,
\href{http://www.arXiv.org/abs/hep-th/9603161}{{\tt hep-th/9603161}}.
%%CITATION = HEP-TH/9603161;%%.

\bibitem{Sen:1996vd}
A.~Sen, ``{F-Theory and Orientifolds},'' {\em Nucl. Phys.} {\bf B475} (1996)
  562--578,
\href{http://www.arXiv.org/abs/hep-th/9605150}{{\tt hep-th/9605150}}.
%%CITATION = HEP-TH/9605150;%%.

\bibitem{Affleck:1983mk}
I.~Affleck, M.~Dine, and N.~Seiberg, ``{Dynamical Supersymmetry Breaking in
  Supersymmetric QCD},'' {\em Nucl.Phys.} {\bf B241} (1984)
493--534.
%%CITATION = NUPHA,B241,493;%%.

\bibitem{Gaberdiel:1997ud}
M.~R. Gaberdiel and B.~Zwiebach, ``{Exceptional Groups from Open Strings},''
  {\em Nucl. Phys.} {\bf B518} (1998) 151--172,
\href{http://www.arXiv.org/abs/hep-th/9709013}{{\tt hep-th/9709013}}.
%%CITATION = HEP-TH/9709013;%%.

\bibitem{DeWolfe:1998zf}
O.~DeWolfe and B.~Zwiebach, ``{String Junctions for Arbitrary Lie Algebra
  Representations},'' {\em Nucl. Phys.} {\bf B541} (1999) 509--565,
\href{http://www.arXiv.org/abs/hep-th/9804210}{{\tt hep-th/9804210}}.
%%CITATION = HEP-TH/9804210;%%.

\bibitem{Mikhailov:1998bx}
A.~Mikhailov, N.~Nekrasov, and S.~Sethi, ``{Geometric Realizations of BPS
  States in ${\mathcal{N}}\!=2$ Theories},'' {\em Nucl. Phys.} {\bf B531}
  (1998) 345--362,
\href{http://www.arXiv.org/abs/hep-th/9803142}{{\tt hep-th/9803142}}.
%%CITATION = HEP-TH/9803142;%%.

\bibitem{GrassiHalversonShaneson:Math}
A.~Grassi, J.~Halverson, and J.~Shaneson, ``{Resolution and Deformation of
  Elliptic Fibrations},''.

\bibitem{GrassiHalversonShaneson:PhysicsCodim2}
A.~Grassi, J.~Halverson, and J.~Shaneson, ``{Matter Representations from
  Deformations of Higher Codimension Singularities},''.

\bibitem{Katz:1996ht}
S.~H. Katz, D.~R. Morrison, and M.~R. Plesser, ``{Enhanced Gauge Symmetry in
  Type II String Theory},'' {\em Nucl.Phys.} {\bf B477} (1996) 105--140,
\href{http://www.arXiv.org/abs/hep-th/9601108}{{\tt hep-th/9601108}}.
%%CITATION = HEP-TH/9601108;%%.

\bibitem{Arnold}
V.~I. Arnol'd, ``NORMAL FORMS OF FUNCTIONS IN NEIGHBOURHOODS OF DEGENERATE
  CRITICAL POINTS,'' {\em Russian Mathematical Surveys} {\bf 29} (1974), no.~2,
  10.

\bibitem{Slodowy}
P.~Slodowy, ``Simple singularities and Complex Reflections,''.

\bibitem{Rossi}
M.~Rossi and L.~Terracini, ``MAPLE subroutines for computing Milnor and Tyurina
  numbers of hypersurface singularities with applications to Arnol'd
  adjacencies,'' \href{http://www.arXiv.org/abs/arxiv:0809.4345}{{\tt
  arxiv:0809.4345}}.

\bibitem{KatzMorrison}
S.~{Katz} and D.~R. {Morrison}, ``{Gorenstein Threefold Singularities with
  Small Resolutions via Invariant Theory for Weyl Groups},''.

\bibitem{Looijenga}
E.~J.~N. Looijenga, {\em Isolated singular points on complete intersections},
  vol.~77 of {\em London Mathematical Society Lecture Note Series}.
\newblock Cambridge University Press, Cambridge, 1984.

\bibitem{Halverson:2013ska}
J.~Halverson, ``{Anomaly Nucleation Constrains SU(2) Gauge Theories},'' {\em
  Phys.Rev.Lett.} {\bf 111} (2013) 261601,
\href{http://www.arXiv.org/abs/1310.1091}{{\tt 1310.1091}}.
%%CITATION = ARXIV:1310.1091;%%.

\bibitem{Dasgupta:1999ss}
K.~Dasgupta, G.~Rajesh, and S.~Sethi, ``{M Theory, Orientifolds and G -
  Flux},'' {\em JHEP} {\bf 9908} (1999) 023,
\href{http://www.arXiv.org/abs/hep-th/9908088}{{\tt hep-th/9908088}}.
%%CITATION = HEP-TH/9908088;%%.

\end{thebibliography}\endgroup

\end{document}